\pdfoutput=1
\documentclass[manuscript,acmlarge]{acmart}

\newcommand{\bheading}[1]{\hfill \break \noindent{\textbf{#1.}}}

\AtBeginDocument{%
  \providecommand\BibTeX{{%
    \normalfont B\kern-0.5em{\scshape i\kern-0.25em b}\kern-0.8em\TeX}}}

\setcopyright{acmlicensed}
\copyrightyear{2024}
\acmYear{2024}
\acmDOI{XXXXXXX.XXXXXXX}

\acmBooktitle{}

\usepackage{graphicx}
\usepackage{subfigure}
\usepackage{soul}
\usepackage[normalem]{ulem} 
\usepackage{xparse} 
\usepackage{xcolor}
\definecolor{darkgreen}{RGB}{60, 150, 50}
\definecolor{darkgrey}{RGB}{180, 180, 180}
\definecolor{darkblue}{RGB}{20, 40, 200}
\usepackage{fancyhdr}
\useunder{\uline}{\ul}{}
\usepackage{multirow}
\usepackage{array}

\newcommand{\cataractbot}{\textit{CataractBot}}
\newcommand{\link}[1]{\textcolor{blue}{#1}}

\newcommand{\imwutcut}[1]{}
\newcommand{\imwutpaste}[1]{\textcolor{black}{#1}}
\newcommand{\imwutadd}[1]{\textcolor{black}{#1}}
\newcommand{\imwutdelete}[1]{}
\newcommand{\imwuthighlight}[1]{\textcolor{black}{#1}}
\newcommand{\imwutaddagain}[1]{\textcolor{black}{#1}}
\newcommand{\imwutdeleteagain}[1]{}

\newcommand{\anonymousHospital}{Sankara Eye Hospital}

\usepackage{setspace} 
\usepackage{etoolbox}

\begin{document}


\title[CataractBot]{CataractBot: An LLM-Powered Expert-in-the-Loop Chatbot for Cataract Patients}

\author{Pragnya Ramjee}
\authornote{Equal contribution.}
\orcid{0000-0003-0061-2624}
\affiliation{%
  \institution{Microsoft Research}
  \city{Bangalore}
  \country{India}
}
\email{t-pramjee@microsoft.com}

\author{Bhuvan Sachdeva}
\orcid{0009-0002-1946-684X}
\authornotemark[1]
\affiliation{%
  \institution{Microsoft Research}
  \city{Bangalore}
  \country{India}}
\email{b-bsachdeva@microsoft.com}

\author{Satvik Golechha}
\orcid{0009-0000-5274-1060}
\affiliation{%
  \institution{Microsoft Research}
  \city{Bangalore}
  \country{India}
}
\email{zsatvik@gmail.com}

\author{Shreyas Kulkarni}
\orcid{0000-0001-8339-2123}
\affiliation{%
 \institution{Microsoft Research}
 \city{Bangalore}
 \country{India}}
 \email{shreyaskulkarni.sak@gmail.com}

\author{Geeta Fulari}
\orcid{0009-0003-2358-810X}
\affiliation{%
  \institution{Sankara Eye Hospital}
  \city{Bangalore}
  \country{India}}
\email{quality@sankaraeye.com}

\author{Kaushik Murali}
\orcid{0000-0002-1385-3227}
\affiliation{%
  \institution{Sankara Eye Hospital}
  \city{Bangalore}
  \country{India}}
\email{kaushik@sankaraeye.com}

\author{Mohit Jain}
\orcid{0000-0002-7106-164X}
\affiliation{%
  \institution{Microsoft Research}
  \city{Bangalore}
  \country{India}}
\email{mohja@microsoft.com}

\renewcommand{\shortauthors}{Ramjee et al.}

\begin{abstract}
  \label{abstract}
The healthcare landscape is evolving, with patients seeking reliable information about their health conditions and available treatment options. 
Despite the abundance of information sources, the digital age overwhelms individuals with excess, often inaccurate information. 
Patients primarily trust medical professionals, highlighting the need for expert-endorsed health information. 
However, increased patient loads on experts has led to reduced communication time, impacting information sharing. 
To address this gap, we developed \cataractbot{}\footnote{Our code is publicly available at 
\link{\url{https://github.com/microsoft/BYOeB}}.}.
\cataractbot{} answers cataract surgery related questions instantly using \imwutadd{an LLM to query} a curated knowledge base, and provides expert-verified responses asynchronously. 
It has multimodal and multilingual capabilities. 
In an in-the-wild deployment study with \imwutdelete{55 participants }\imwutadd{49 patients and attendants, 4 doctors, and 2 patient coordinators}, \cataractbot{} \imwutdelete{proved valuable }\imwutadd{demonstrated potential}, providing anytime accessibility, saving time, accommodating diverse literacy levels, alleviating power differences, and adding a privacy layer between patients and doctors. 
Users reported that their trust in the system was established through expert verification. 
Broadly, our results could inform future work on \imwutdeleteagain{designing }expert-mediated LLM bots.
\end{abstract}




\maketitle

\section{Introduction}
\label{introduction}
The evolving landscape of healthcare witnesses a significant shift with patients assuming more proactive roles in their care journeys~\cite{infowork-cancer-chi10, info-need-before-consult},
resulting in an increased demand for information.
Patients and their caregivers seek accessible, comprehensive, and reliable information about their symptoms, diagnoses, treatment options, potential risks, and preventive measures~\cite{info-needs-review}.
To satisfy these information needs, patients actively explore diverse sources, including online resources, friends and family, support groups, and direct communication with healthcare professionals~\cite{social-cancer-chi10}.
In particular, for major treatments like surgery, satisfying the demand for information becomes crucial~\cite{info-needs-review}, with questions spanning pre-, during-, and post-treatment phases.
Studies highlight the anxiety patients and caregivers experience regarding such treatments~\cite{unanchored-information-cancer-klasnja-2010}, emphasizing the correlation between anxiety and negative clinical outcomes~\cite{anxiety-outcome, anxiety-outcome2}. 
Similarly, access to information has been found to significantly reduce anxiety~\cite{information-display-chi2010} and improve patient satisfaction and clinical outcomes~\cite{doctor-patient-comm-outcome, information-display-chi2010}, thus underlining its pivotal role.

Despite the abundance of available information sources, the digital age often overwhelms individuals with an excess of information, much of which is inaccurate or unreliable~\cite{info-need-internet-review, info-need-before-consult}.
This poses a challenge in discerning trustworthy sources from misleading ones.
Patients also frequently encounter difficulties finding information online at an appropriate level, ranging from oversimplified to overly technical~\cite{powering-ai-chatbot-xiao-2023}.
As a result, patients tend to rely on the experiences of friends and family members who have undergone similar treatments, although identifying them can be challenging~\cite{patient-expertise-locator-civan-2010}.
Therefore, patients primarily trust their doctor and the hospital staff responsible for their treatment, turning to them for any medical or logistical queries~\cite{info-needs-pre-post, info-need-internet-review}.
Studies examining patients' information needs reveal a strong desire for doctor-endorsed health information~\cite{informing-patients}.

However, the escalating pressure on doctors to accommodate more patients has led to reduced time per patient, affecting communication and information sharing.
In developing nations, a lower doctor-to-patient ratio hinders personalized attention and comprehensive guidance~\cite{ai-doctor-rural-wang-2021}, while in developed countries, managing Electronic Health Records (EHR) often diminishes direct patient communication quality and available time~\cite{ehr-conflict-cajander-2019}.
Previous studies~\cite{Tongue2005CommunicationSF, info-need-before-consult} have found that doctors often underestimate patients' information needs and overestimate the amount of information they provide.
However, providing all potentially relevant information risks overwhelming patients with cognitive overload~\cite{info-needs-review}.
Surveys consistently highlight patients' desire for improved communication with their healthcare providers~\cite{Kalamazoo}.
This information exchange extends beyond doctor-patient interactions to encompass relationships between various medical professionals and patients, such as patient-nurse interactions~\cite{nurse-patient-comm, ding-wechat-chi2020}.
Moreover, enhanced access to information not only benefits patients but also has the potential to reduce unnecessary visits, alleviating the burden on doctors and medical staff.
To summarize, as \citet{patient-missing-link} correctly stated: ``\textit{Patients have little access to information and knowledge that can help them participate in, let alone guide, their own care... A simple, non-urgent exchange of questions and answers is often all that is required.}''.

To address patients' information needs, particularly during surgical treatments, we propose a chatbot solution \imwutadd{integrated with medical professionals} to provide \imwutadd{accessible}, reliable and comprehensive responses.
Leveraging the capabilities of large-language models (LLMs) and the widespread use of smartphones and instant messaging services such as WhatsApp, this chatbot serves as a
24/7 resource capable of understanding intricate human queries and providing accurate information.
In collaboration with a tertiary eye hospital in Bangalore, India, we exemplified this experts-in-the-loop chatbot approach by developing \cataractbot{},
which answers queries related to cataract surgery from patients and their attendants.
Recognizing the limitations of generic LLMs in the medical domain, we utilized Retrieval-Augmented Generation~\imwutadd{\cite{rag-lewis-2020}} with an LLM over a custom knowledge base \imwutaddagain{(Section~\ref{systemdesign:responsegeneration})} curated by doctors and hospital staff to provide hospital-specific and culturally sensitive responses.
\cataractbot{} has several features \imwutadd{to enhance adoption and sustained use.
Capitalizing on the widespread adoption of smartphones, we chose a mobile-first design paradigm, deploying the chatbot entirely on the ubiquitous WhatsApp platform.
To cater to diverse hospital visitors with varying literacy levels and technological proficiency, we made \cataractbot{} multilingual (supporting five languages, including English, Hindi, and Kannada) and multimodal (accepting both speech and text inputs)}.
\imwutcut{\imwutdelete{multimodal support} (accepting both speech and text inputs), \imwutdelete{multilingual capabilities} (supporting five languages, including English, Hindi, and Kannada)}
\imwutadd{We use an LLM to process complex and ill-formed queries, generating instant responses from the curated knowledge base.
Crucially, \cataractbot{} incorporates} a system where the LLM's responses are \imwutadd{asynchronously} verified by experts \imwutadd{and corrected if needed}. 
Ophthalmologists verify answers to medical queries and patient coordinators to logistical ones.
These expert-provided edits are used to update the knowledge base, to minimize future expert intervention.
\imwutdelete{\cataractbot{} integrates both synchronous and asynchronous messaging by providing an LLM-generated answer instantly, and subsequently providing verification at the experts' convenience.}

Our work aims to answer the following research questions:
(\textbf{RQ1}) What is the role of \cataractbot{} in meeting the information needs of patients (and attendants) undergoing surgery?
(\textbf{RQ2}) 
How do the different features of \cataractbot{}, including \imwutadd{LLM-generated answers, }experts-in-the-loop and multimodality, contribute to its usage?
(\textbf{RQ3}) How can \cataractbot{} be integrated into the current doctor-patient workflow?
To answer these questions, we conducted an in-the-wild deployment study involving \imwutadd{multiple stakeholders (See Table \ref{tab:stakeholder-roles}):} 49 information seekers\imwutdelete{\footnote{\imwutdelete{We use `participants' or `information seekers' to refer to patients and their attendants, and `experts' to refer to doctors and coordinators.}}} (19 patients, 30 attendants)
and 6 experts (4 doctors and 2 patient coordinators).
Our findings \imwutdelete{revealed }\imwutadd{indicated} that patients and attendants appreciated \cataractbot{} because it alleviated hesitations associated with power differences in asking questions, and accommodated individuals with low literacy and tech proficiency through multilingual and multimodal support.
Patients and attendants reported that their trust in the system was established through the verification performed by doctors and coordinators.
Experts praised the bot as a facilitator, introducing a privacy layer between them and patients, and providing flexibility through asynchronous communication.\imwutdelete{~Moreover, experts noted improved responses over time, likely due to continuous updates to the knowledge base.}

Our main contributions are:
(1) The design and development of a novel \imwutadd{LLM-powered,} experts-in-the-loop\imwutadd{, multilingual, multimodal} chatbot,
created in collaboration with doctors and hospital staff to assist cataract surgery patients with their information needs.
(2) The deployment of our bot in a real-world \imwutadd{multi-stakeholder} setting, among 49 cataract surgery patients/attendants, four doctors, and two patient coordinators at a tertiary eye hospital in India.
(3) Drawing from a comprehensive mixed-methods analysis of gathered data, the paper offers \imwutdeleteagain{pivotal }insights specific to \cataractbot{} and, \imwutdeleteagain{broader implications of }\imwutaddagain{more broadly,} LLM-powered experts-in-the-loop chatbots.


\section{Related Work}
\label{relatedwork}

Our work is primarily informed by two areas of relevant research: solutions addressing patients' information needs and the use of conversational agents, particularly those powered by LLMs, in healthcare settings.

\subsection{Patients' Information Needs}
\label{relatedwork:infoneeds}
Patients and their attendants seek  information throughout the various stages of their medical journey, such as preparing for doctor consultations~\cite{radionova2023impacts, awe-chi18}, verifying diagnoses and treatment plans~\cite{eriksson2021patients, bickmore2009taking, info-needs-review}, assessing the need for clinical interventions~\cite{yang2023talk2care, migrant-worker-chatbots-tseng-2023}, and aiding in the recovery process~\cite{wang-2021-social-support, patient-expertise-locator-civan-2010, fadhil2018conversational}.
Studies emphasize that satisfying these diverse information needs improves patient understanding, motivates adherence to treatment regimens, and contributes to overall satisfaction with healthcare services~\cite{informing-patients}.
Patients primarily trust doctors with their medical queries~\cite{ding2020gettinghealthcare}.
However, traditional in-person consultations pose difficulties, including high costs~\cite{bodenheimer2013primary}, long wait times, and a strong hierarchy in the patient-physician relationship~\cite{fochsen2006imbalance}.
Patients often miss opportunities to ask additional (clarification) questions during these brief encounters, as they face information overload~\cite{preanesthetic-info-ferre-2020, info-need-before-consult}. 
Doctors also struggle with time due to staff shortages straining healthcare ecosystems~\cite{bodenheimer2013primary, akerkar2004doctor}.
Thus, patients and caregivers are increasingly turning to the internet for easier access to medical information~\cite{powering-ai-chatbot-xiao-2023, chandwani2016whosthedoctor}.
However, the challenge lies in identifying relevant and appropriate information.
Individuals struggle to understand complex health information or may even consume inaccurate information, making the process frustrating~\cite{zhang2010contextualizing,arora2008frustrated,clarke-health-info-needs-barriers-2016}.

Information seekers may also consult experienced patients within their social networks who can share insights related to managing similar health conditions~\cite{patient-expertise-locator-civan-2010, social-cancer-chi10}.
In the absence of such known individuals, people may resort to social media, which is not only ineffective for quality health support due to widespread misinformation and poor management~\cite{malki2024wildheadline}, but also risky in terms of personal privacy~\cite{choudhury2014seeking}.
\citet{chandwani2016whosthedoctor} found that although most doctors regard the practice of seeking health information online as an affront to their authority, some see it as inevitable--and even as an opportunity to elevate patients' health literacy through technology.

Recent work in the field of Computer-Human Interaction has explored novel avenues to address these medical information gaps.
For instance, \citet{awe-chi18} 
propose interactive `science museum' exhibits in clinical waiting rooms to educate children and their parents about sickle cell anemia,
\citet{information-display-chi2010} and \citet{bickmore2009taking} propose information displays in hospital rooms to provide patients with real-time status updates and treatment details, and \citet{phone-info-gap-chi12} propose using mobile phones to present dynamic, interactive reports on patients' progress and care plans throughout their emergency department stay.
Although these are specific solutions for certain diseases and infrastructures, discussions have been ongoing around making Electronic Health Record (EHR) data accessible to patients~\cite{pa-ehr-chi19, patient-missing-link, chung2017supporting}.
Patient-facing EHR services have received mixed reviews, with healthcare professionals expressing concerns about adding strain to already understaffed healthcare systems, while patients generally welcome the idea~\cite{pa-ehr-chi19}.

\imwutadd{
Prior work has also explored ``chat'' as an interface to address patients' information needs, by connecting them with doctors in both synchronous~\cite{textmssg-brenna-chi23} and asynchronous~\cite{johansson2020general, johansson2020patients} manners.
Synchronous messaging requires both patient and doctor to be simultaneously active, allowing instant feedback and coherent dialogue.
On the other hand, asynchronous messaging platforms like WhatsApp~\cite{yadav-feeding-2019}, WeChat~\cite{ding-wechat-chi2020}, SMS~\cite{perrier2015pregnantkenyanwomen}, email~\cite{makarem2016email}, or web portals \cite{mamykina2008scaffolding, tuli2018menstrupedia} enable communication at the convenience of both parties.
Research indicates that both doctors and patients prefer asynchronous texts over instant audio/video calls~\cite{stamenova2020uptake, johansson2020general}, despite lacking instant feedback~\cite{textmssg-brenna-chi23}.
}

In summary, patients need reliable, relevant, and timely information.
However, obtaining this from busy doctors and hospital staff is challenging.
This dichotomy results in patients either receiving generic, unverified, and hard-to-understand information from the internet and social circles, or, if lucky, obtaining personalized, verified answers from healthcare experts---at a higher cost.
Our solution aims to strike a balance by providing a generic, high-quality LLM answer, later verified by a medical professional with minimal increase to their workload.

\subsection{\imwutadd{LLM} Chatbots in Healthcare}
\label{relatedwork:chatbotshealthcare}
Chatbots enable natural language conversations, offering a minimal learning curve and a personalized experience\imwutadd{~\cite{thirunavukarasu2023llmsinmedicine}}\imwutdelete{~\cite{farmchat-imwut18, jo2023understanding, roller2020recipes, preanesthetic-info-ferre-2020, chung2017supporting, desai2023okgoogle}}.
\imwutadd{AI-powered chatbots surpass rule-based platforms in language understanding, }allowing users to build complex queries message by message while retaining context~\cite{powering-ai-chatbot-xiao-2023}.
\imwutadd{With the advent of LLMs, building complex chatbots has become significantly easier, enabling} researchers to apply \imwutadd{LLM-powered} bots across diverse healthcare settings\imwutadd{~\cite{thirunavukarasu2023llmsinmedicine}}\imwutdelete{~\cite{wutz2023factors}}, including \imwutdelete{scheduling appointments~\cite{ittarat2023personalized}, }pre-consultation data collection~\cite{li2024preconsolutationchatbots}, \imwutdelete{assisting new mothers~\cite{yadav-feeding-2019}, }chronic disease management~\cite{montagna2023data,hao2024LLMapplicationsinpatienteducation}, 
and mental health support~\cite{kumar2023exploring, jo2023understanding,kim2024mindfuldiary}. 
Commercial solutions, such as Babylon Health~\cite{babylon}\imwutdelete{, Ada, and Florence,\footnote{\imwutdelete{Babylon Health: https://www.emed.com/uk. Ada: https://ada.com/. Florence: https://florence.chat/.}}}, now provide individuals with instant access to any health-related information.
\imwutdelete{All of these bots are fully automated, with some even using LLMs. }\imwutpaste{However, LLM-powered healthcare tools have generally been criticized for hallucinations, errors, and a lack of transparency in their reasoning~\cite{au2023ai, denecke2024potentialofllms}.}
As the involvement of doctors in these bots is\imwutadd{--at most--}limited to the data curation step~\cite{powering-ai-chatbot-xiao-2023}, they are susceptible to generating inaccurate information and are ``\textit{not ready for clinical use}''~\cite{au2023ai}.

\imwutcut{LLMs have been criticized for hallucinating and not displaying the reason behind their outputs~\cite{fogel2018artificial}.}
Several methods have been proposed to address these issues, including chain-of-thought prompting~\cite{chainofthought}, retrieval-augmented generation~\cite{rag-lewis-2020}, and few-shot learning.
\citet{seitz2022trust} identified that patients' trust in healthcare chatbots is primarily based on a rational evaluation of their capabilities, while trust in human medical professionals relies more on emotional connections and personal rapport. 
Integrating doctors into the loop with automated \imwutadd{LLM} chatbots could combine the best of both worlds, a concept we propose in \cataractbot{}.
\imwutpaste{In our work, we leverage LLM-based responses in real-time, while human medical experts verify and respond asynchronously.
This hybrid approach is a novel and reliable method for addressing healthcare information needs.}

\imwutadd{
We chose cataract surgery--one of the world's most common procedures~\cite{mcghee2020cataractcommon}--as the use case for developing and testing our system. 
This decision was motivated by our longstanding collaboration with \anonymousHospital{}, an eye hospital in India.
Also, cataract surgery involves well-defined protocols and common patient queries, making it an ideal starting point for testing the bot's ability to deliver accurate, standardized information from a curated knowledge base.
However, our solution is versatile, and we believe it can effectively support a range of health conditions,
including mental health and chronic care.}
Given our specific focus on cataract surgery, we examined literature at the intersection of ophthalmology and LLMs.
\citet{bernstein2023comparison} reported that responses generated by LLMs to patient eye-care-related questions are comparable to those written by ophthalmologists. Based on similar findings, \citet{ittarat2023personalized} proposed integrating chatbots into ophthalmology practices to provide 24/7 support for common inquiries on eye conditions.
Additionally, \citet{yilmaz2024talking} reported that chatbots are an effective tool for educating cataract surgery patients.
Our work contributes to this body of literature.

Finally, real-world healthcare deployments of AI are limited due to socio-technical uncertainties in the last mile~\cite{andersen2021realizing}. 
To avoid such shortcomings, as recommended by \citet{thieme2023foundation, Thieme2023designingmental}, we designed and studied our system within a tertiary eye hospital in India and 
examined its integration into clinical workflows.

\section{Formative Study}
\label{formativestudy}
The development of \cataractbot{} was informed by our literature survey, our understanding of the Indian healthcare ecosystem, and insights from a formative study conducted  at \anonymousHospital{}.
To identify requirements,
we conducted semi-structured interviews with 4 ophthalmologists and 2 patient coordinators between July-Aug 2023.
\imwutadd{We solely interviewed experts, as we aimed to understand the status quo, including their workflows, the types of questions patients ask, and experts' concerns regarding patients' knowledge gaps.}
\imwutadd{We then conducted pilot tests with all end users--patients, attendants, doctors, and coordinators--which iteratively informed the design of \cataractbot{}.}
Interview participants had an average age of 38.2$\pm$8.3 years, with 11.3$\pm$7.3 years of work experience \imwutadd{(Table~\ref{tab:demography-experts} in Appendix \ref{appendix:participant demography})}.
They participated in the study voluntarily without compensation.
The interviews were conducted in English and lasted around 45 minutes each.
All sessions were audio-recorded and transcribed by the interviewer.
\imwutadd{The interviewer open coded the interview transcripts, following inductive thematic analysis~\cite{braun2006thematicanalysis} to identify key design requirements for the chatbot.
Two researchers met regularly to review the emerging codes and finalized the high-level themes.}

\subsection{\imwutadd{Background}}
\label{formative-study-background}
\imwutadd{Our design requirements were shaped by the workflows and information needs specific to cataract surgery at \anonymousHospital{}.
In the current workflow, patients (with their attendants) first consult a doctor who diagnoses the cataract's maturity and type, recommending the urgency of surgery and potential complications.
After deciding on surgery, the patient and attendant meet with a patient coordinator for guidance on pre- and post-operative measures, lens options across different budget ranges, and logistical tasks such as insurance and scheduling.
Patients also undergo pre-operative medical tests.
For further questions, they can either visit the hospital to meet the doctor or contact the coordinator by phone.
On the day of surgery, patients arrive with their attendants in the morning, undergo surgery by mid-morning, and are discharged by the afternoon or evening.
The doctor and coordinator provide post-surgery care instructions, covering medication, exercise, bathing, travel, and screen time.
One week post-surgery, patients return for a follow-up appointment with the operating doctor.}

\subsection{Design Requirements}
\label{formative-study-design-requirements}
\bheading{Accuracy}
\label{formative-study-design-requirements-accuracy}
The chatbot's ability to answer both medical and logistical questions with accuracy, precision and contextual relevance was found to be essential given the medical context.
While a standard LLM like GPT-4~\cite{gpt4_report} might offer accurate information, it may not be tailored to India- or hospital-specific nuances.
For instance, doctors informed us that although the phacoemulsification method is prevalent globally for cataract surgeries, they often favor manual small incision-based procedures~\cite{singh2017review} due to their cost-effectiveness.
Similarly, cultural considerations significantly influence pre- and post-surgery questions.
As a doctor stated, ``\textit{People have unique questions... like `when can I do
\imwutdelete{Ayurvedic }nasal yoga post-op?'.}''
Further, queries related to diet require awareness of Indian food options for precise guidance.
Thus, our bot relies on a custom knowledge base curated by hospital staff.
\imwutcut{To ensure accurate and contextual responses, our chatbot relies on a knowledge base curated by hospital medical staff, including doctors, patient coordinators, and members of Quality Control and Patient Safety team.}

\bheading{Trustworthiness}
Access to trustworthy information is crucial during medical treatment.
As one doctor noted, ``\textit{The success of treatment is not only in medicine, but the trust they place on us.}''
Patients and their attendants typically place their utmost trust in the operating doctor
and hospital staff.
Even when patients learn something from the internet or other sources, they often confirm that information with their doctor.\imwutdelete{~E.g., a doctor noted, ``\textit{They come and ask us... my friends have told me this... is it correct or not?}''}
To ensure trustworthiness, we decided that (a) each response would be verified and, if necessary, edited by a doctor or patient coordinator, and (b) patients (or their attendants) would be explicitly notified upon expert verification.
Note: We intentionally designed \cataractbot{} to be non-specific to individual patients by not connecting it with the hospital's scheduling system or EHR data.
This decision was made because LLMs can memorize training data, including personally identifiable information (PII) like emails and phone numbers, and leak it during inference~\cite{borkar2023can}, raising privacy concerns and potentially breaking users' trust.

\bheading{Timeliness}
Discussions with hospital staff highlighted the crucial need for real-time responses.
Patient coordinators noted that patients frequently reach out to them over multiple phone calls to ask logistical questions and relay medical queries to their consulting doctors.
Due to experts' busy schedules, immediate responses are not always feasible.
This often results in patient seeking information from other sources (including the internet), which can propagate misinformation, or fail to completely address their information needs.
To counter this, we decided that the chatbot should provide an instantaneous response to every patient query using the LLM and a custom knowledge base.
Subsequently, these responses get reviewed and refined by doctors/coordinators.
Moreover, doctors mentioned their unavailability for several hours while performing surgeries.
To ensure swift verification, \cataractbot{} forwards queries to an `escalation' expert if they remains unverified for 3 hours.

\bheading{Usability}
After cataract surgery, patients can reference their discharge summaries for any queries, but this
is challenging due to their temporarily reduced vision.
Prior research~\cite{gronvall2013beyondselfmonitoring} has identified the crucial role family members play in monitoring the patient's health. 
Coupled with doctors' input (``\textit{Patients ask hardly 10\% of the question... attendants are the ones who ask.}''), this underscored the necessity for the bot to be usable by
both the patients and their attendants.
To cater to diverse backgrounds, 
we made the chatbot multilingual and multimodal.
We chose WhatsApp as the messaging platform due to its widespread use in India~\cite{statista2022} and therefore minimal learning curve and quick onboarding without additional installations, despite the richer feature set of other platforms like Telegram or the development flexibility of a custom chatbot.
Considering doctors' demanding schedules, often involving consulting $\sim$50 patients and performing $\sim$10 cataract surgeries a day, we implemented features to minimize their bot-related workload: (1) one-click interaction to verify an answer, (2) allow experts to provide corrections using informal messages~\cite{beasley2009short}, (3) handle spelling and grammar errors, and (4) use expert edits to enhance the bot's knowledge base, minimizing similar edits in future.

\section{\cataractbot{} System Design}
\label{systemdesign}

Here, we describe the key components of the \cataractbot{} system (Figure~\ref{fig:cataractbotsystem})--input language and modality, response generation and verification, escalation and reminders--and provide a detailed account of its implementation.

\begin{figure*}
  \includegraphics[width=\textwidth]{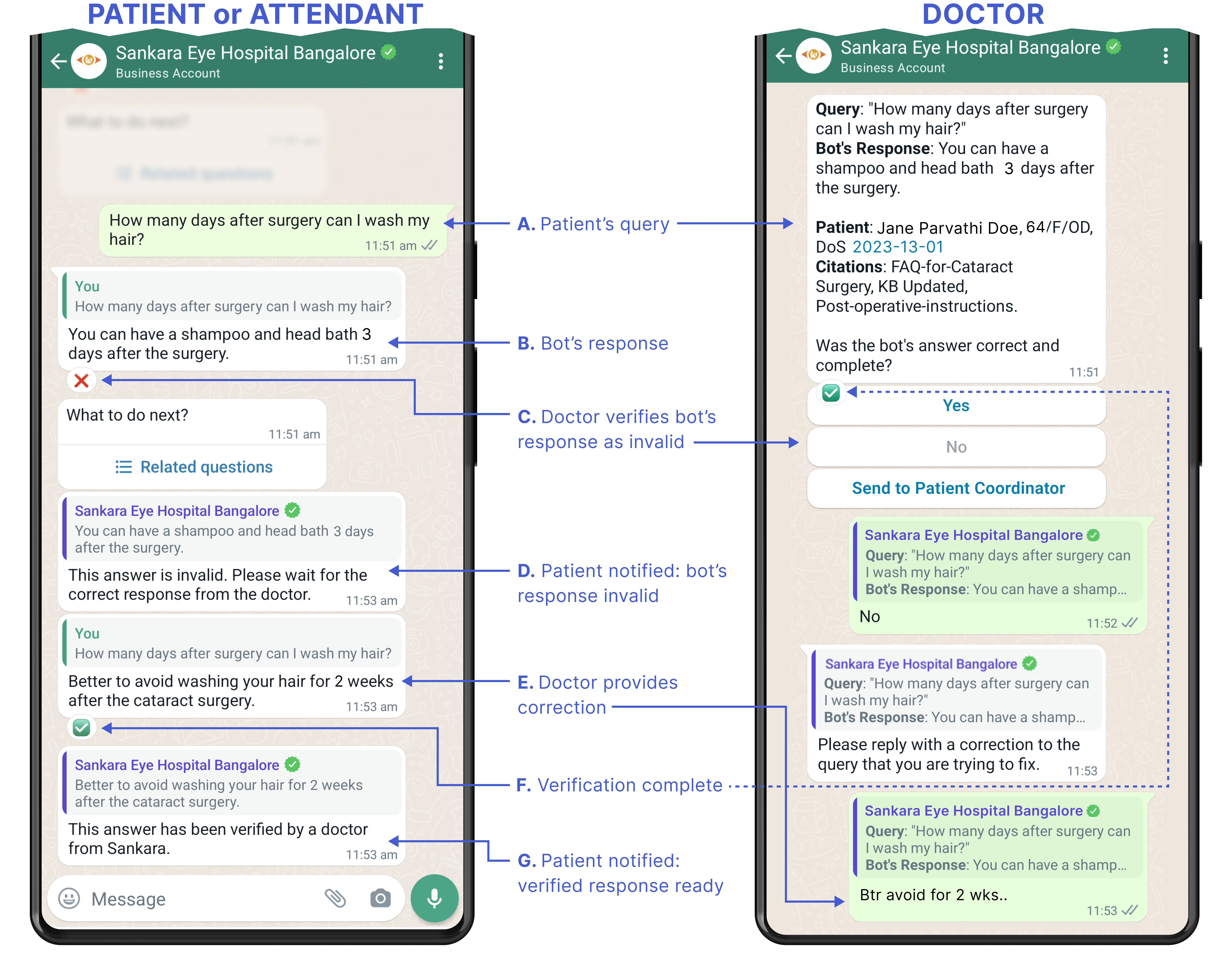}
  \caption{\cataractbot{} provides an initial response to the patient/attendant by querying the knowledge base. The doctor (or coordinator, if the question is logistical) verifies and corrects this response, and the patient/attendant is notified.}
  \Description{CataractBot is displayed with a split-screen view showing both patient/attendant (left) and doctor (right) perspectives. The workflow demonstrates: 1) A patient asks "How many days after surgery can I wash my hair?"; 2) CataractBot initially responds that shampooing is allowed 3 days after surgery; 3) The doctor reviews this response, knowing it is intended for patient Jane Parvathi Doe (64/F/OD) who had cataract surgery on 2023-12-01, and marks it as incorrect; 4) The doctor provides the correct guidance, without having to worry about spelling or grammar, "Btr avoid for 2 wks.."; 5) The patient is first notified that the bot's answer was invalid and to wait for the doctor; and 6) Finally, the patient receives the doctor-verified corrected message stating "Better to avoid washing your hair for 2 weeks after the cataract surgery."}
  \label{fig:cataractbot_textmessage}
\end{figure*}

\begin{table}[]
\small
\centering
\imwutadd{
\caption{\imwutadd{Stakeholders and their roles in the \cataractbot{} socio-technical system}}
\label{tab:stakeholder-roles}
\begin{tabular}{lll|l}
\hline
\multicolumn{3}{l|}{\textbf{Stakeholder}} &
  \textbf{Role} \\ \hline
\multicolumn{1}{l|}{\multirow{2}{*}{\begin{tabular}[c]{@{}l@{}}Information \\ seeker\end{tabular}}} &
  \multicolumn{2}{l|}{Patient} &
  \begin{tabular}[c]{@{}l@{}}Person scheduled for cataract surgery.\\ Asks \cataractbot{} surgery-related questions.\end{tabular} \\ \cline{2-4} 
\multicolumn{1}{l|}{} &
  \multicolumn{2}{l|}{Attendant} &
  \begin{tabular}[c]{@{}l@{}}Person accompanying the patient (e.g., child of the patient).\\ Asks \cataractbot{} surgery-related questions.\end{tabular} \\ \hline
\multicolumn{1}{l|}{\multirow{5}{*}{Expert}} &
  \multicolumn{1}{l|}{\multirow{3}{*}{Doctor}} &
  \begin{tabular}[c]{@{}l@{}}Operating \\ doctor\end{tabular} &
  \begin{tabular}[c]{@{}l@{}}Surgeon scheduled to operate on the patient.\\ Verifies \cataractbot{}’s answers to users’ medical questions.\end{tabular} \\ \cline{3-4} 
\multicolumn{1}{l|}{} &
  \multicolumn{1}{l|}{} &
  \begin{tabular}[c]{@{}l@{}}Escalation \\ doctor\end{tabular} &
  \begin{tabular}[c]{@{}l@{}}Senior surgeon. Verifies \cataractbot{}’s answers to medical \\ questions that the operating doctor is unable to address in time.\end{tabular} \\ \cline{3-4} 
\multicolumn{1}{l|}{} &
  \multicolumn{1}{l|}{} &
  \begin{tabular}[c]{@{}l@{}}Knowledge \\ base expert\end{tabular} &
  \begin{tabular}[c]{@{}l@{}}Senior surgeon. Selects and edits verified answers for \\ addition to \cataractbot{}'s knowledge base.\end{tabular} \\ \cline{2-4} 
\multicolumn{1}{l|}{} &
  \multicolumn{1}{l|}{\multirow{2}{*}{\begin{tabular}[c]{@{}l@{}}Patient \\ coordinator\end{tabular}}} &
  \begin{tabular}[c]{@{}l@{}}Operating \\ coordinator\end{tabular} &
  \begin{tabular}[c]{@{}l@{}}Liaison between patients/attendants and operating doctors. \\ Verifies \cataractbot{}’s answers to users' logistical questions.\end{tabular} \\ \cline{3-4} 
\multicolumn{1}{l|}{} &
  \multicolumn{1}{l|}{} &
  \begin{tabular}[c]{@{}l@{}}Escalation \\ coordinator\end{tabular} &
  \begin{tabular}[c]{@{}l@{}}Senior coordinator. Verifies \cataractbot{}’s answers to logistical \\ questions that the operating coordinator is unable to address in time.\end{tabular} \\ \hline
\end{tabular}%
}
\end{table}

\subsection{Components}
\imwutdelete{Below, we describe the key components of our \cataractbot{} system.}

\subsubsection{Input Language and Modality}
\anonymousHospital{} in Bangalore (in the state of Karnataka), referred to as the Silicon Valley of India, caters to patients from various linguistic, educational, and technical backgrounds, including those from the Information Technology sector and neighboring states.
Analysis of a 2011 census highlights Bangalore as one of India's most linguistically diverse cities~\cite{census2011}.
To accommodate this varying patient demography, \cataractbot{} is designed to support five languages: English, Hindi (the local language of nine states in India), Kannada (the local language of Karnataka), and Tamil and Telugu (the local languages of two neighboring states of Karnataka).
As the expert may not be proficient in all five languages, their interactions with the bot are exclusively in English.
Upon onboarding, the language preference of the patient and their attendant is collected through an online form, and \cataractbot{} initiates the conversation by sending a set of `welcome messages' in the chosen language.
Among these messages is an option allowing users to modify their preferred language in future.
To engage with users with different levels of literacy 
\cataractbot{} supports both text and speech inputs.
For every voice message, \cataractbot{} sends both a text response and a spoken version of the text as an audio message.

Recognizing the challenge users face in initiating a conversation, (`Blank Page Syndrome'~\cite{blank-page-syndrome, pwr}), \cataractbot{} provides a set of three frequently asked questions along with the welcome message.
Additionally, every response includes a ``\textit{What to do next?}'' prompt, offering three related questions based on the preceding query (Figure~\ref{fig:cataractbot_audiomessage}F in Appendix \ref{appendix:multimodal}).
Users can select one of the suggested questions or pose a different query to continue the conversation.
This feature lowers the barrier to entry (as formulating questions has been found to be non-trivial for low-literate individuals~\cite{migrant-worker-chatbots-tseng-2023}), enhances the natural flow of conversation, and saves time.

\imwutcut{
\begin{figure*}
  \includegraphics[width=\textwidth]{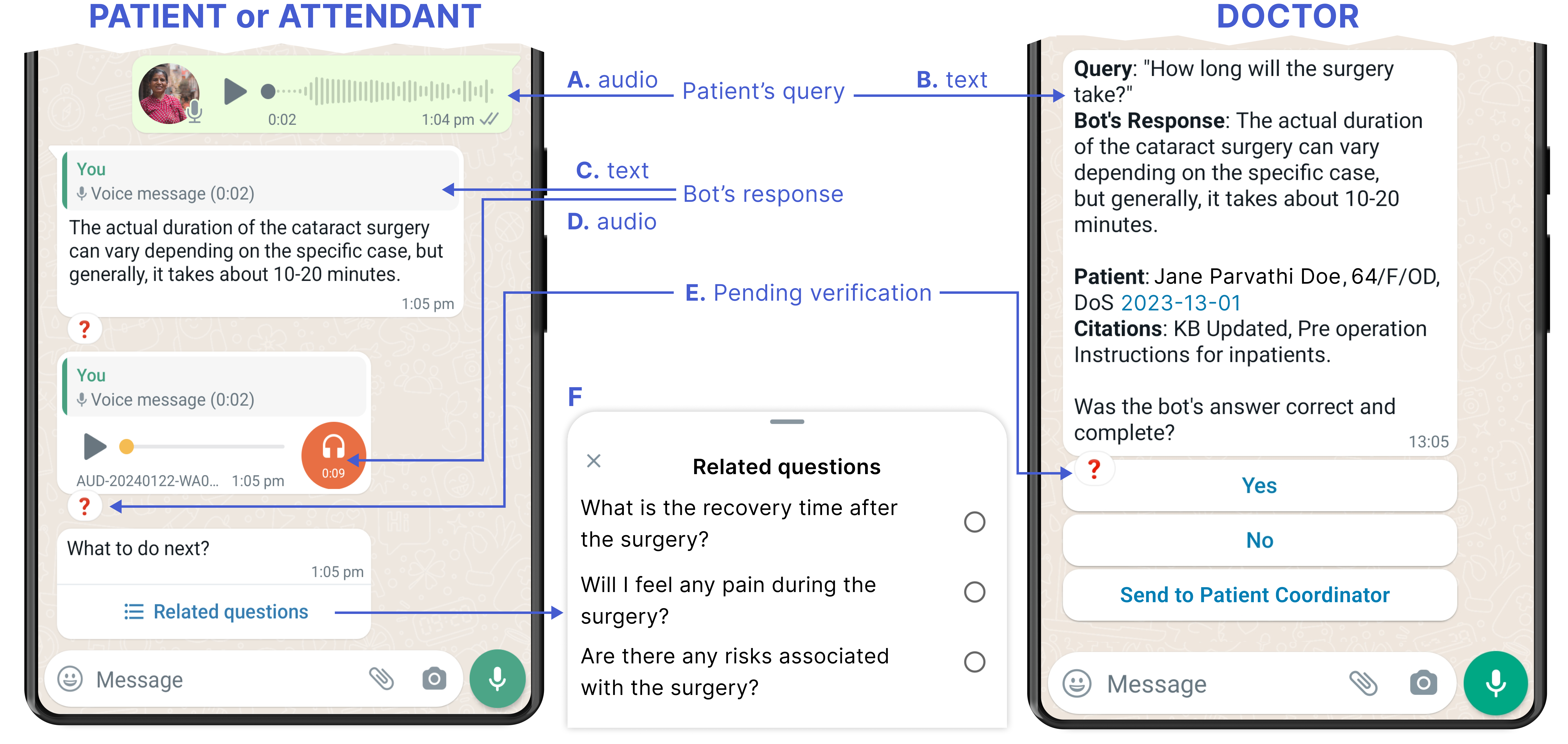}
  \caption{\imwutcut{A question asked (using audio), receiving an unverified response and Related Questions from \cataractbot{}.}}
  \Description{Asking a question to \cataractbot{} through audio}
  \label{fig:cataractbot_audiomessage_cut}
\end{figure*}
}

\subsubsection{Response Generation, Verification and Icons}
\label{systemdesign:responsegeneration}

Upon receiving a message, \cataractbot{} classifies it as a medical question (e.g., dos and donts before and after surgery), logistical question (e.g., scheduling or insurance related), small talk (e.g., greetings or expressions of gratitude), or `other', and provides a response in real-time.
For medical and logistical questions, the bot strictly employs the knowledge base curated by the hospital's medical team to generate an appropriate response.
\imwutadd{This custom knowledge base comprises of various cataract surgery and hospital-specific documents curated by hospital staff (doctors, patient coordinators, and members of Quality Control and the Patient Safety team).
These documents, comprising approximately 30 pages in total, include the Consent Form, Standard Operating Procedures, FAQs, Pre- and Post-Operative Guidelines, and Hospital Information.
Some documents, like Post-Operative Guidelines, were included in patients' discharge summaries.
Others, such as SOPs and FAQs, existed in the hospital ecosystem but were not directly accessible to patients.}
\imwutaddagain{The knowledge base also included verified question–answer pairs that were added over the course of the deployment. 
We provide details of this incremental updating process in Section~\ref{systemddesign:knowledgebaseupdate}.}
Grounded in this custom knowledge base, the bot-generated response includes a question mark icon \includegraphics[height=0.65\baselineskip]{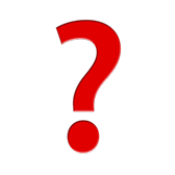} as a `reaction' indicating the unverified status (Figure \ref{fig:cataractbot_audiomessage}E in Appendix \ref{appendix:multimodal}).
In instances where the knowledge base lacks an answer, the bot responds with a template ``\textit{I don't know}'' response.
For small talk messages (such as ``\textit{Hello}'' or ``\textit{Thank you for the information}''), the chatbot provides corresponding small talk responses.

For medical questions, the patient's operating doctor (Table \ref{tab:stakeholder-roles}) receives a message comprising of the question asked, the bot's response, and patient demographics (Figure \ref{fig:cataractbot_textmessage}A).
As \citet{rajashekar2024humanalgorithmic} found that citations improve trust in LLM-generated responses, this message also includes citations of the documents used to generate the response.
The doctor is prompted with the question, ``\textit{Is the answer accurate and complete?}'' offering three response options: ``\textsf{Yes}'', ``\textsf{No}'', and ``\textsf{Send to Patient Coordinator}''.
Tapping `\textsf{Yes}' replaces the question mark icon \includegraphics[height=0.65\baselineskip]{Figure/Red_QuestionMark.pdf} in the information seeker's received answer with a green tick icon \includegraphics[height=0.65\baselineskip]{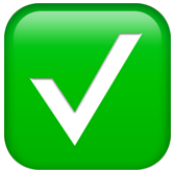}, confirming verification (Figure \ref{fig:cataractbot_textmessage}F).
Additionally, the bot notifies the information seeker that the answer has been verified, tagging that particular response (Figure \ref{fig:cataractbot_textmessage}G). 
Tapping `\textsf{No}' replaces the question mark \includegraphics[height=0.65\baselineskip]{Figure/Red_QuestionMark.pdf} with a red cross icon \includegraphics[height=0.65\baselineskip]{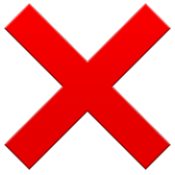}, indicating an incorrect answer (Figure \ref{fig:cataractbot_textmessage}C), and asks the information seeker to await a corrected response from the doctor (Figure \ref{fig:cataractbot_textmessage}D).
The doctor is asked to provide a correction; they are not required to edit the bot's message, instead, they can offer a correction in free form text.
The system combines the bot's initial answer with the doctor-provided correction to frame a new response, which is delivered to the user with a green tick icon \includegraphics[height=0.65\baselineskip]{Figure/Green_CheckMark.pdf} (Figure \ref{fig:cataractbot_textmessage}E).
This new response is not sent to the doctor, to minimize their workload by limiting verification to a single round.
If a question is misclassified, i.e., a logistical question is sent to the doctor, they have the option to `\textsf{Send to Patient Coordinator}'.
Similar to the doctors' workflow, patient coordinators verify and provide corrections to the bot's responses for logistical questions.
We deliberated between displaying unverified responses in real-time or only verified responses post-expert verification and correction.
Through our formative study and discussions with hospital staff, it became clear that providing real-time responses was crucial.
Delayed responses might lead to patients making repeated calls to the hospital or resorting to online sources, failing to meet their information needs.

\subsubsection{Escalation Mechanism}
The \cataractbot{} system employs an intricate escalation and reminder mechanism to ensure swift verification.
In the doctor's WhatsApp interface, unanswered questions are marked with a question mark icon \includegraphics[height=0.65\baselineskip]{Figure/Red_QuestionMark.pdf} (Figure \ref{fig:cataractbot_audiomessage}E in Appendix \ref{appendix:multimodal}) and answered questions with a green tick \includegraphics[height=0.65\baselineskip]{Figure/Green_CheckMark.pdf}, enabling them to easily identify pending queries.
If an answer is not verified by the operating doctor within three hours, it is automatically sent to the designated escalation doctor (Table \ref{tab:stakeholder-roles}).
Both the operating and escalation doctors can then verify/correct the response.
Selecting `Yes' or `Send to Patient Coordinator' immediately marks the query with a green tick (Figure \ref{fig:cataractbot_textmessage}F) for both the doctors, indicating task completion.
However, if either doctor selects `No', the green tick appears only after that doctor provides a correction.
If neither the operating nor escalation doctor verifies a question within six hours, both receive a reminder notification indicating the pending status.
Additionally, every four hours during working hours (at 8 am, 12 pm, and 4 pm), a list of all questions pending verification for more than six hours is sent to both the operating and escalation doctors.
This reminder was added based on the doctors' feedback.
This workflow is mirrored for the operating and escalation patient coordinators.

\subsubsection{Knowledge Base Update Process}
\label{systemddesign:knowledgebaseupdate}
To minimize experts' labor, we use expert-provided edits to update the knowledge base, which increases the likelihood of `Yes' responses from experts to similar questions in the future.
However, certain responses, such as those specific to individual patients (\textit{e.g.}, ``\textit{When to reach the hospital for my surgery?}'') must not be added to the knowledge base as they are not generalizable.
We appointed a senior cataract surgeon as the `knowledge base expert' (Table \ref{tab:stakeholder-roles}), who received a spreadsheet via email at 8pm daily.
Their task involves reviewing each row, which consists of question-answer pairs, and determining whether the information should be added to the knowledge base (by responding with a `Yes'/`No' in the `Should Update Knowledge Base?' column), and if so, modifying as needed the `Final Answer for Knowledge Base' column, containing the bot's updated answer based on the expert's correction.
At 3 am daily, the system extracts `Question' and `Final Answer for Knowledge Base' data from all rows marked `Yes', and append these to the knowledge base in the `expert-FAQ' document.
Additionally, we prompt the LLM to prioritize this `expert-FAQ' document within the knowledge base.
This ensures that \cataractbot{} improves with these updates, gradually enhancing its accuracy over time.

\subsection{Implementation Details}
\label{systemdesign:implementation}
\begin{figure*}
  \includegraphics[width=\textwidth]{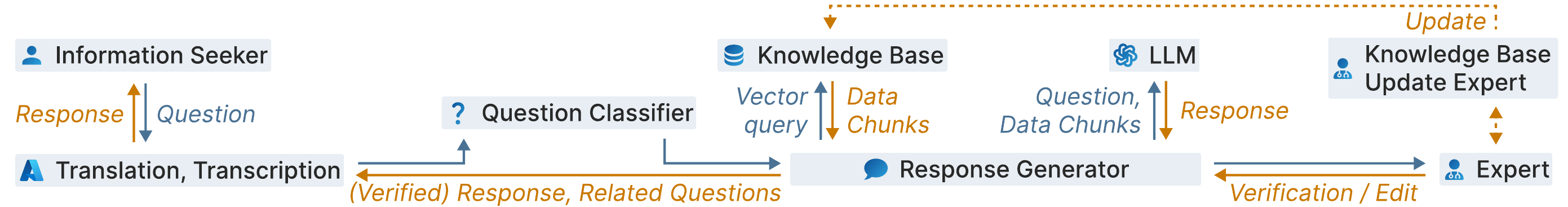}
  \caption{Overview of \cataractbot{} System}
  \Description{This image shows a workflow diagram for a CataractBot. The process begins with an Information Seeker on the left who submits a Question, which goes through a Question Classifier. The classified question then queries a Knowledge Base, which outputs Data Chunks that feed into an LLM, which outputs an (unverified) response, shared with an Expert for verification or edits. The resulting output is sent back to the Information Seeker as a Verified Response with Related Questions. Knowledge Base Update Experts can use the Expert edits to provide Updates to the Knowledge Base.}
  \label{fig:cataractbotsystem}
\end{figure*}

The \cataractbot{} system relies on these five components (Figure~\ref{fig:cataractbotsystem}): 
(a) LLM: for response generation, 
(b) Vector Database: for storing and retrieving the custom knowledge base, 
(c) Language Technologies: for translation and transcription, 
(d) WhatsApp Services: for message exchange, and 
(e) Cloud Storage: for storing interaction logs. 

We opted for GPT-4, the leading LLM at the time of our system development in May 2023~\cite{gpt4_report}.
\imwutcut{The custom knowledge base comprises of various cataract surgery and hospital-specific documents, including Consent Form, SOP (Standard Operating Procedure), FAQs, Pre- and Post-Operative Guidelines, and Hospital Information.}
The documents for the custom knowledge base are ingested as data chunks into a Chroma vector database, using the OpenAI model `text-embedding-ada-002' to generate embeddings.
Upon receiving a question, GPT-4 classifies the query type. 
For a medical/logistical question, \cataractbot{} employs a retrieval-augmented generation~\cite{rag-lewis-2020} approach.
This involves performing a vector search on the knowledge base to extract the three most relevant data chunks.
GPT-4 is then prompted (full prompt available in Appendix~\ref{appendix:prompt}) to generate an answer for the query from these data chunks.
If the answer is not present, the bot responds with an ``\textit{I don't know}'' message.

The prompt underwent several iterations.
To validate its effectiveness, a doctor and a patient coordinator recorded 153 questions posed by patients/attendants scheduled for cataract surgery over a week.
We used our GPT-4 prompt to obtain answers to these questions from the custom knowledge base, and subsequently the generated responses were evaluated by the same doctor and coordinator.
This not only helped refine the LLM prompt, but also led the medical staff to improve existing documents and contribute additional documents to enrich the knowledge base.
This process was iterated thrice.

Additionally, we utilize GPT-4 to incorporate corrections provided by doctors/coordinators to generate the final expert corrected response.
This involved prompting GPT-4 with the patient's question, the initial response from the bot, and the expert-provided correction, and asking it to generate a revised answer considering the expert's input.
Our experience showed that GPT-4 executed this task well.
Regarding language integration, GPT-4 primarily comprehends and responds in English~\cite{mega-eval}. 
To bridge the language gap, our system adhered to the standard approach of translating Indic languages (Hindi, Kannada, Tamil, and Telugu) into English for input using Azure AI Translator~\cite{aitranslator}.
It then translated GPT-4's English responses back into the respective Indic languages for output.
To facilitate speech as input and audio as output, \cataractbot{} uses Azure speech-to-text and text-to-speech models.
(Note: GPT-4 lacks support for speech input.)
Finally, all text and audio interactions between the \cataractbot{} system and users (including patients, attendants, operating and escalation doctors, coordinators, and knowledge base expert) get logged in a cloud storage for further analysis.

\section{Study Design}
\label{studydesign}
To evaluate the effectiveness of this LLM-powered experts-in-the-loop chatbot in addressing the informational needs of patients undergoing cataract surgery, we conducted a mixed-method user study during Nov 2023-Jan 2024 at our collaborators' institute, \anonymousHospital{}.
This is one of the leading tertiary eye care and teaching institutions in Bangalore, India.
It typically attends to more than 500 patients every day, and conducts over 50 cataract surgeries daily.
This study was approved by both the Scientific and Ethics Committees of \anonymousHospital{}.
None of the study participants---patients, attendants, doctors, or coordinators---received any financial incentives for their participation, in accordance with hospital norms.

\subsection{Procedure for Patients and Attendants}
\label{studydesign:procedureforinfoseekers}
\imwutcut{As per the hospital's protocol, once a patient opts for cataract surgery based on their doctor's recommendation, the patient and their attendant needs to meet a patient coordinator.
The coordinator helps with general guidance on pre- and post-operative measures, scheduling of the surgery date, and logistical arrangements.
Toward the end of this interaction,}
The operating patient coordinator was tasked to assess the patient's eligibility for the study based on specific criteria:
(a) age of 18 or older,
(b) fluent in one of the five languages supported by \cataractbot{},
(c) scheduled for cataract surgery within a week,
(d) no history of cataract surgery in the last 6 months (as recent patients would likely have minimal informational needs), and
(e) having one of the three participating operating doctors as their surgeon.
If these criteria were met, the patient and their attendant were directed by the coordinator to meet a researcher (the first author) stationed at the hospital.

The researcher introduced them to \cataractbot{} and outlined the study's protocol which involved using the bot for around two weeks and engaging in an interview pre- and post-surgery.
If they agreed to participate, participants were requested to sign a consent form, after which the researcher filled a web-based onboarding form.
This form included details such as the patient and attendant's phone numbers linked to WhatsApp, preferred languages, consulting doctor and coordinator, surgery date, and basic demographic information (age, gender, and education).
Upon form submission, participants received `welcome messages' from \cataractbot{}.
The researcher instructed them to ask a trial question, either by choosing from suggested questions or by typing/speaking in their preferred language.
Upon receiving a response, participants were briefed on the icons and expert verification system. Throughout this process, the researcher encouraged them to ask any questions regarding the bot's usage.
After onboarding, participants received two daily reminder messages (at 7:30 am and 4 pm) prompting them to ``\textsf{ask any cataract surgery related questions}'' until one week post surgery, when their access was revoked.

We conducted two semi-structured interviews: one on the surgery day and another a week post-surgery.
These specific days were chosen because they coincided with the patient's required hospital visits.
The interviews explored their overall experience, specific features they liked or disliked, suggestions for improvement, and questions regarding the trustworthiness, timeliness, and accuracy of responses.
Both interviews followed a similar structure, differing only in how the bot supported the participant before versus after surgery.
After each interview, participants were requested to fill out a chatbot usability form.
Those who agreed, rated their chatbot experience through eight questions, using a five-point Likert scale, covering aspects like ease of use, understandability, and timeliness  (Appendix \ref{appendix:usabilityform}). 
These questions were adapted from metrics outlined in prior usability evaluations of healthcare bots~\cite{abd2020technical, holmes2023validatingscale}.
Patient-attendant pairs were interviewed together but filled separate usability forms.

All interviews with patients and attendants were conducted by the first author in English, Hindi, Kannada, or Tamil, as the researcher is fluent in these languages.
(Note: Participants who selected Telugu as their preferred language during onboarding were interviewed in one of the other four languages.)
Interviews on the day of surgery took place in person at the hospital, while post-surgery interviews were conducted either in person at the hospital or through pre-scheduled telephone calls.
This decision was influenced by the substantial waiting time on the surgery day---patients typically spend about five hours in the hospital, with the procedure itself taking $\sim$45 minutes.
Conversely, during the post-surgery visit, patients receive priority and experience minimal waiting time.
Each interview typically spanned 20 to 40 minutes and was audio-recorded with the participant's consent.

\subsection{Procedure for Doctors and Coordinators}
We recruited four doctors and two patient coordinators.
Three doctors served as operating doctors, and one as both an escalation doctor and knowledge base expert.
Both patient coordinators specialize in handling cataract surgery patients, with one serving as a coordinator and the other as an escalation coordinator.
The escalation doctor and escalation coordinator held senior positions in the hospital and carried administrative responsibilities.
All participating experts attended one of two one-hour training sessions conducted by three researchers.
During these sessions, \cataractbot{} was demonstrated, and experts were onboarded. 
The researchers assumed patient roles, asking medical and logistical questions for the doctors and coordinators to verify and edit. 
The first session revealed a few bugs, and feedback was collected to enhance the bot.
For example, a doctor suggested displaying the question first (rather than the patient's demographics) in verification messages, as only the top portion is visible in `tagged' WhatsApp messages (Figure \ref{fig:cataractbot_textmessage}E).
After using \cataractbot{} for two weeks or longer, the participating experts were interviewed by the first author.
These interviews followed a semi-structured format, focusing on the bot's usability, its integration into their workflow, and its impact on interactions with patients.
The interviews were audio-recorded, conducted in English, in-person, and typically lasted 45-60 minutes.

\subsection{Participants}
\label{studydesign:participants}
A total of 31 patient-attendant pairs (Table~\ref{tab:demography-patients} in Appendix \ref{appendix:participant demography}), consisting of 19 patients and 30 attendants, took part in our study.
Due to the predominantly elderly demographic undergoing cataract surgery, 12 patients lacked access to WhatsApp/smartphones and were unable to participate but their attendants used the bot. 
One patient was not accompanied by an attendant.
The average age of the 19 patients was 58.8$\pm$8.01 years. 
Although females comprised 18 (32.7\%) of all participants, only 3 (15.8\%) were patients. 
This aligns with prior work showing a gender disparity in accessing cataract surgery in India \cite{prasad2020gender, ye2020female}.
Patients had diverse preferred languages (5 English, 5 Hindi, 5 Kannada, 3 Tamil, and 1 Telugu) and education levels (6 $\leq$Grade 10, 4 Grade 12, 6 Bachelors, and 3 Masters).
In contrast, the attendants were younger, predominantly fluent in English, and well-educated.
The average age of the 30 attendants (11 female) was 37.2$\pm$10.3 years.
A majority (23) preferred English, with 2 each favoring Hindi, Kannada, and Tamil, and 1 preferring Telugu. 
Education levels were mostly high, with 10 having Masters and 13 having Bachelors degrees.
Among the 31 patient-attendant pairs, 22 participated in the surgery-day interview, and 10 took part in the one-week post-surgery interview.
Various factors influenced participation, including time constraints (resulting in a few declined interviews) and data reaching saturation.
Additionally, four doctors
and two patient coordinators 
participated in the study (Table~\ref{tab:demography-experts} in Appendix \ref{appendix:participant demography}). The roles of escalation doctor and knowledge base expert were performed by the same doctor. 
Note: there was a partial overlap with two doctors and one coordinator participating in both this study and the formative study.

\subsection{Data collection and analysis}
\label{studydesign:dataanalysis}
We performed a mixed-method analysis to systematically analyze the collected data comprising of usage logs, interview transcripts, and usability survey responses. 
\cataractbot{} interaction logs and usability survey data were quantitatively analyzed.
Descriptive statistics and statistical tests (such as t-tests and ANOVAs) were used to analyze the count and type of questions asked, verification styles and response times of doctors and coordinators, edits performed by the knowledge base update expert, and responses to the usability survey.
Note: For ANOVAs, the sphericity assumption was tested using Mauchly's test, and in case of sphericity violations, the Greenhouse-Geisser correction was applied.
The first author translated and transcribed all interviews (totalling 11.5 hours) into English soon after they were conducted.
Throughout the study, all authors regularly engaged in discussions to review observations and interviews.
For interview transcripts, we conducted an inductive thematic analysis~\cite{braun2006thematicanalysis} approach, with the first author open-coding all interviews line-by-line.
Subsequently, two authors collaboratively reviewed these codes to identify interesting themes in the data (similar to \cite{okolo2021cannot}).
We iterated on these themes to distill higher-level themes (like ``Usage'' and ``Features''), that we present in our findings below.
\imwuthighlight{Note: While there were a total of 49 patients and attendants using \cataractbot{}, at times, we treat them as 31 participants (patient-attendant pairs).
This choice stems from instances highlighted during interviews, where patients instructed attendants to query the bot using either of their phones.
Additionally, both were interviewed together.
Only when comparing their distinct roles, we analyze their usage log data separately.}

\subsection{Positionality}
All authors are of Indian origin, and currently reside in Bangalore, India.
One author is a practising ophthalmologist, performing over 10 cataract surgeries weekly, and another author is part of the Quality Control and Patient Safety team at the same hospital.
Both were involved in designing the study, shaping research questions, providing feedback on initial bot versions, curating the knowledge base, and ensuring that any risk of harm to the patients was mitigated.
The ophthalmologist also served as the knowledge base expert.
Two authors specialize in HCI and design, three possess healthcare research experience, and four have expertise in software development.
We approached this research drawing on our individual experiences and learning from working at the intersection of computing and healthcare in India.
Our interest in making healthcare more accessible to the masses has informed the design of \cataractbot{} and guided our study design.
\section{Findings}
\label{findings}
Overall, 31 patient-attendant pairs sent 343 messages to \cataractbot{} (11.06$\pm$8.44 messages/pair).
225 messages were medical questions, 87 logistical, 27 small talk, and 4 others.
\imwutpaste{A researcher manually classified all questions into 12 categories, identified in a bottom-up manner (similar to \cite{textmssg-brenna-chi23, clarke-health-info-needs-barriers-2016}).
Several questions fell into multiple categories.
Figure \ref{fig:cataractbot_CHARTS}A illustrates the relative significance of each category before, on, and after the day of surgery.}
\imwutpaste{Before the surgery, questions focused on the procedure (like ``\textsf{anaesthesia}'', ``\textsf{lens}'', ``\textsf{safety}'') and admission/discharge logistics\imwutdelete{~(e.g., ``\textsf{eat or not before the surgery}'', ``\textsf{documents to bring for surgery}'', and ``\textsf{take blood thinners on the day of surgery}'')}, while post-surgery questions centered on medication and recovery (e.g., ``\textsf{washing hair}'', ``\textsf{doing yoga}''\imwutdelete{~and ``\textsf{watching TV}''})}.
\imwutdelete{One day before the surgery, questions related to the ``\textsf{schedule}'' peaked, as information seekers awaited calls from patient coordinators about their surgery time.}
Doctors directly approved (said `\textsf{Yes}' to) 69.8\% (157) of the bot-generated answers 
while patient coordinators directly approved 58.6\% (51).
\imwutadd{17 of the 22} (77.3\%) usability survey respondents found \cataractbot{}'s responses to be useful, appropriate, and informative, while \imwutadd{20} (91.0\%) indicated willingness to use it in the future.

Below,\imwutdelete{~in our findings, we delve into how  participants used \cataractbot{} (answering RQ1 and RQ3), examine the impact of novel features introduced in \cataractbot{} (answering RQ2), and explore privacy and trust implications of using an LLM-powered chatbot (answering RQ1).}
\imwutadd{
Section \ref{findings:rq1} presents the reasons behind patients' and attendants' engagement with the bot (answering RQ1).
Section \ref{findings:rq2} examines the impact of key features of the bot, including the LLM, experts-in-the-loop
and support for multiple stakeholders, languages and modalities (RQ2).
Finally, Section~\ref{findings:rq3} describes the integration of \cataractbot{} into the doctor-patient workflow and discusses issues of personalization and accountability (RQ3).
Table~\ref{tab:findings-summary} in Appendix~\ref{appendix:findings-summary} presents a summary of key findings across sections.}

\subsection{\imwutadd{Bot's Role in Addressing Information Needs}}
\label{findings:rq1}
\imwutdelete{\subsection*{6.1 \cataractbot{} Usage}}

\subsubsection{Reasons for (Not) Using \cataractbot{}}
\label{findings:rq1:reasons}
Information seekers used \cataractbot{} to address questions they had forgotten or were uncomfortable to ask the doctor or patient coordinator, clarifying answers, and seeking updates.
D1 stated that most information seekers come prepared for face-to-face conversations, optimizing their limited time and avoiding multiple visits.
\cataractbot{} can alleviate the burden of having to remember every question.
Ten information seekers highlighted that the bot helped recall information provided 
during in-person consultations, which they had either forgotten or not completely understood.\imwutdelete{~As A13 stated: ``\textit{We are humans... cannot pay complete attention all the time.}'' }
Patients found comfort in ``\textit{written}'' information, 
available as a reference (similar to \cite{farmchat-imwut18}), ``\textit{I need not remember everything.}'' (A20).
Six participants also mentioned using the bot to ``\textit{double-check}'' (A7) and seek ``\textit{reassurance}'' (A17), despite already knowing certain answers.

Due to existing power differences between doctors and patients~\cite{knowledge-power-williams-2014} and the judgmental attitudes of some doctors, individuals often hesitate to ask questions.
Experts acknowledged that educated individuals 
are comfortable posing questions. 
In contrast, ``\textit{less educated patients who don't speak English}'' are the ones for whom \cataractbot{} ``\textit{can do wonders}'' (D3).
Five participants, all with a Grade-12 education or below, relied heavily on \cataractbot{} asking 57 questions in total, as 
they reported being unaccustomed to seeking information online. 
One of them expressed that checking with the bot was more comfortable than approaching a doctor, citing the bot's non-judgmental nature: ``\textit{(The patient) has cholesterol issues and heart surgery... the doctor would make faces when I ask about food intake... but not the bot.}'' (A4).

Despite these reasons, 
8 out of 49 information seekers \imwutdelete{did not use the bot, }\imwutadd{used the bot minimally, sending only one or two messages}.
Two key reasons emerged.
First, six participants mentioned that their in-person interactions with the doctor and patient coordinator sufficed to address their queries.
This suggests that \cataractbot{} could play a more pivotal role in primary healthcare centers, where the human infrastructure may be less capable of addressing information needs or where power imbalances make individuals less likely to seek answers.
Second, two participants considered cataract surgery to be a ``\textit{simple}'' procedure, leading them to have no questions.
This indicates that a similar bot could be valuable for more complex (surgical) procedures.

\subsubsection{\imwutadd{``\textsf{I don't know}'' Responses.}}
\label{findings:rq1:idk}
Participants complained about receiving ``\textsf{I don't know}'' responses (57 total)
A total of 36 (10.5\%) messages requested status updates---seeking information about the surgery time and discharge time. 
Due to privacy concerns, \cataractbot{} was not integrated with the hospital's patient management system, resulting in ``\textsf{I don't know}'' responses.
Patient coordinators later provided answers.
While participants acknowledged the appropriateness of ``\textsf{I don't know}'' responses, 
they suggested providing a ``\textit{tentative response}'' (A10) if possible.
For instance, 
to the question ``\textsf{Mine is general anesthesia or local anesthesia?}'' 
A10 expected an answer such as ``\textit{99\% of cataract surgeries at \anonymousHospital{} are done under local anesthesia}''.
From an expert's perspective, responding to certain ``\textsf{I don't know}'' questions was challenging as they lacked the full context.
E.g., A21 asked:
`\textsf{I want to postpone the surgery to Feb... is this okay or do I need to get this operated immediately?}'', receiving an ``\textsf{I don't know}'' response.
To correct this, the doctor said: ``\textsf{Please come and check}''.
This correction frustrated the attendant, who stated: ``\textit{But (the patient) already went through the complete eye check; the doctor know her condition... I don't want to come and visit the doctor [again]... it will waste 2-3 hours.}'' (A21).

\imwutpaste{
When participants received an ``\textsf{I don't know}'' response, they sought information from other sources, particularly the internet.
For time-sensitive questions, participants reported using the internet to ``\textit{double-check}'' the initial LLM-generated, unverified response due to delays in receiving an expert verified response.
Four participants, accustomed to relying on the internet for their informational needs, reserved \cataractbot{} exclusively for questions requiring expert opinions, such as ``\textsf{When can i do exercises as anulom vilom?}'' (P27).
According to D4, without \cataractbot{}, patients might have resorted to the internet and risked following inaccurate advice, given the unavailability of doctors for immediate consultations. 
\cataractbot{} improves the likelihood of receiving accurate responses, as it uses a custom knowledge base.}
\begin{figure*}
  \includegraphics[width=\textwidth]{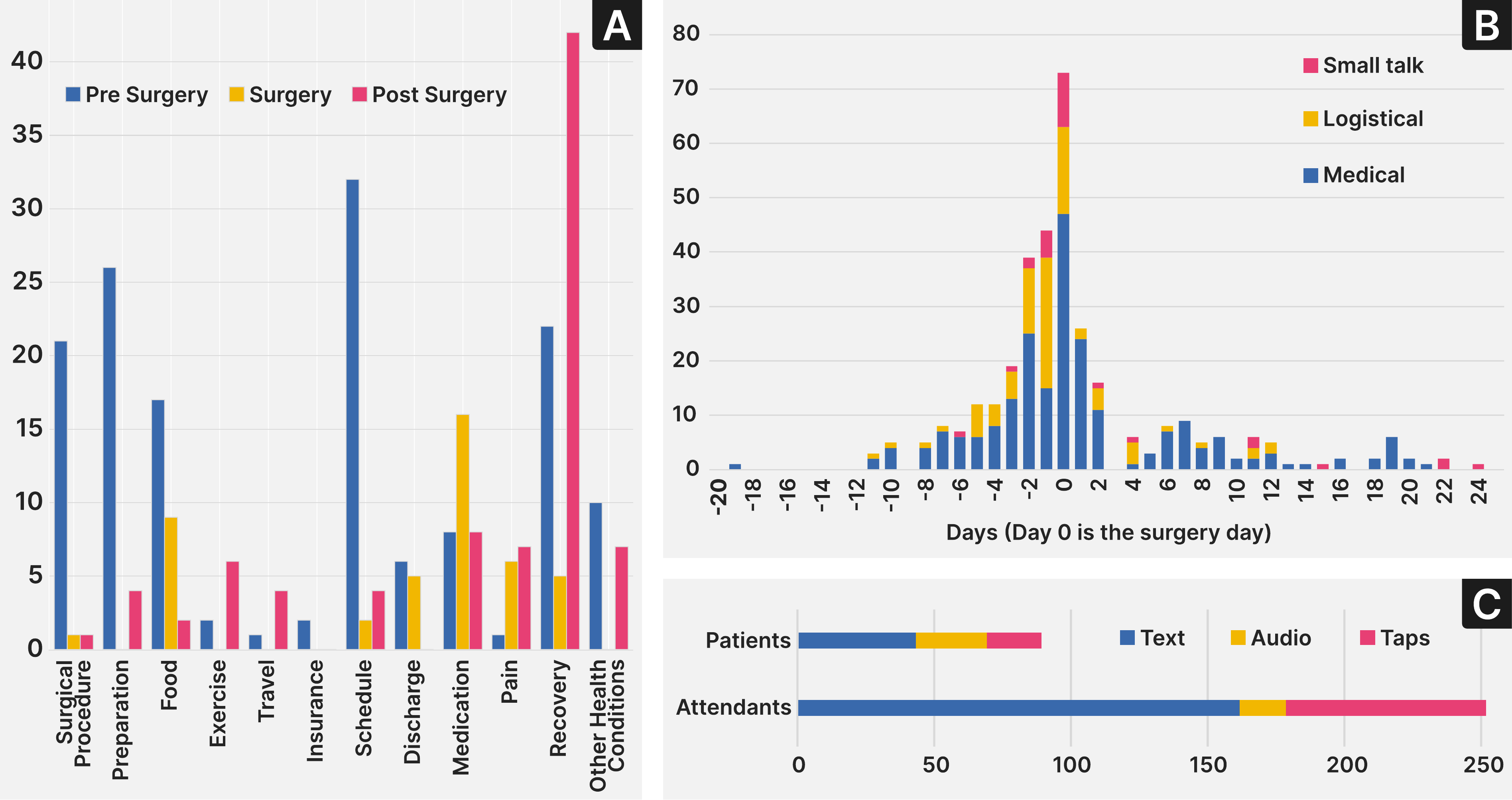}
  \caption{\textbf{A.} (Left) Messages sent per day pre- and post-surgery. \textbf{B.} (Top right) Messages sent before surgery, on the day of surgery, and after surgery. \textbf{C.} (Bottom right) Message input modality.}
  \Description{Figure A (left) displays message categories sent pre-surgery, during surgery, and post-surgery, with medical procedure questions dominating pre-surgery, medication and pain concerns prominent during surgery, and recovery questions increasing post-surgery. Figure B (top right) shows message frequency relative to surgery day (Day 0), revealing a sharp communication spike in the days immediately before and after surgery, with peak activity occurring on surgery day and decreasing gradually afterward. Figure C (bottom right) compares communication methods between patients and attendants, showing patients primarily use text messaging while attendants more frequently utilize a combination of text, audio, and taps.}
  \label{fig:cataractbot_CHARTS}
\end{figure*}

\imwutdelete{\\\textit{6.1.3 Question Type and Style.}}
\label{findings:questiontypeandstyle}

\subsection{\imwutadd{Effect of Key Features on the Bot's Usage}}
\label{findings:rq2}

\subsubsection{\imwutadd{LLM-powered}}
\label{findings:features:llmpowered}
\imwutpaste{Patients and attendants found the LLM-generated responses from the curated knowledge base were useful, }\imwutadd{and experts described them as ``\textit{highly accurate}'', noting that they minimized their workload for corrections.}
\imwutpaste{These instant responses saved time as information seekers did not have to constantly wait, either physically or over a call. 
Seven participants emphasized that traveling is time-consuming, and contacting busy hospitals over the phone is challenging due to high patient loads, as ``\textit{the phone is always busy}'' (A4).}
\imwutpaste{Further, the robustness of the GPT-4 model enabled \cataractbot{} to appropriately handle complex queries from participants with limited literacy, who struggled to formulate questions, consistent with prior research~\cite{farmchat-imwut18}.
For example, P18 asked ``\textsf{Tomorrow operation When should come No phone call received yet?}''.
\cataractbot{} was able to address 45/65 such questions, of which 26 were asked using audio in an Indic language.
}
\imwutpaste{
Typing in Indic languages using the Latin script (\textit{e.g.,} instead of Devanagari script for Hindi) was rare, occurring in only three messages.
\imwutdelete{This was not supported by \cataractbot{} due to current translation technology limitations, and it once responded with ``\textsf{You seem to be asking the question in a language different from the selected language...}''.}
In \imwutdelete{the other }two instances, the messages mixed Hindi and English, and the bot responded accurately:
\begin{quote}
    \textbf{A8}: \textsf{agar opration k baad pain ho raha to kya karna hai?}\\
    \textbf{Bot}: \textsf{If the patient is experiencing pain post-surgery, they should report to the doctor immediately...}
\end{quote}}

\imwutadd{For experts, the LLM enabled them to swiftly correct the bot's responses using natural language, without worrying about grammatical or spelling errors.
For example, in response to a query, D2 added the correction: ``\textsf{Btr to book an appt}''. 
In other instances, experts also instructed the LLM to remove specific sentences, relying on it to generate appropriate responses from their feedback.
}

\subsubsection{Experts-in-the-loop}
\label{findings:features:expertinloop}
Information seekers appreciated the novel feature of having human experts verify and update the auto-generated responses (Figure \ref{fig:cataractbot_textmessage}F), serving as the key reason for their trust in and engagement with the bot. 
For instance, ``\textit{That's why I gave my WhatsApp} (number to onboard the bot)\textit{, so I can talk directly to the doctor.}'' (A2).
\imwutdelete{
Six information seekers explicitly mentioned trusting the bot as a system because of their trust in the hospital, where they had chosen to undergo surgery.
As \cataractbot{} was deployed through \anonymousHospital{}'s verified Facebook Business account, its default WhatsApp username included the hospital's name and displayed a green tick indicating `verified' status (Figures \ref{fig:cataractbot_textmessage}, \ref{fig:cataractbot_audiomessage}).
Moreover, the integration of the bot with the hospital was reinforced by the patient coordinator advising them to register and use the bot.}
\imwutpaste{
A12 added: 
``\textit{The trust only came when I saw the green tick mark... Before that, I was also in the question mark zone... thinking  it's a machine... But then somebody is checking it, confirming it... It's actually trustworthy then.}''}
Despite our concerns that the icons might be complex, participants found them self-explanatory:
\begin{quote}
    ``It's simple... Even my dad understood. I didn't tell him [about] the tick or question mark... He told me, ``\textit{Doctor verified... the tick mark came}''... If MY dad was able to use it, I think anyone can.'' -- A13.
\end{quote}

\imwutdelete{The experts-in-the-loop ensured that the bot's response were more reliable and ``\textit{not fake}'' (P18).
This is in contrast to a few participants who complained about receiving false information from Google search results.}
\imwutpaste{Although the experts-in-the-loop feature increased trust in the final responses, it had a negative impact on perceived bot intelligence.
Four participants noted that when an expert corrected \cataractbot{}'s initial answer, their trust in the bot's unverified answers reduced.
For instance,
``\textit{After some time, when the doctor said it is invalid, I was like, okay, should I even trust the bot?}'' (P19).
In total, 12.8\% (44) of the bot's generated answers (excluding ``\textsf{I don't know}'') for medical and logistical queries were marked incorrect by the expert.
This could be due to the limited custom knowledge base, the experts' high standards for tone and structure of answers, or our prompting strategy favoring caution.}
Also, since participants knew that these messages were being reviewed by experts, four used the bot to send direct messages to them instead of asking questions (e.g., ``\textsf{Dr. Give me a call?}'' (P19)).

\imwutpaste{Experts appreciated the intermediary role of the bot, introducing a layer of separation from patients: ``\textit{From a doctor's standpoint, it is very good because it puts in a curtain between us.}'' (D4).
}
Further analysis revealed that their suggested edits fell into five categories: adding missing information (76.8\%), asking clarification questions (14.6\%), making factual corrections (11.0\%), asking the patient to visit the hospital (8.8\%), and removing unnecessary information (3.7\%).
The length of expert corrections in characters varied significantly (t(91)=5.2, p<0.05) with 152.4$\pm$104.9 for medical questions and 50.9$\pm$42.6 for logistical questions.

Finally, experts did not have access to the final answer sent to information seekers after their suggested edits.
This limited verification to a single round, and was designed to minimize their workload.
However, this lack of transparency sometimes left experts uncertain if \cataractbot{} had understood their corrections. 
Only the knowledge base update expert had access to the updated answer sent to the information seeker.
\subsubsection{Multimodality}
\label{findings:features:inputmode}
\imwutdelete{Of the 343 messages sent by information seekers, 206 (60.1\%) were text, 43 (12.5\%) audio and 94 (27.4\%) related questions.}
The 19 patients sent 89 messages (43 text, 26 audio, and 20 taps), while the 30 attendants sent 254 messages (163 text, 17 audio, and 74 taps) (Figure \ref{fig:cataractbot_CHARTS}C).
One-way ANOVAs on message types found statistical significance for attendants (F(2, 87)=13.74, p<.001), but not for patients (F(2, 54)=0.77, p=0.46).
Post-hoc pairwise comparisons, corrected with Holm's sequential Bonferroni procedure, indicated statistically significant differences between `text' and `audio' (t(29)=4.8, p<.01) and `text' and `taps' (t(29)=3.0, p<0.01) for attendants.
The minimal usage of audio by attendants could be due to demographic differences, with the majority being under 45 years and well-educated.
Audio messages were predominantly used by older semi-literate patients.
For instance, ``\textit{I can't read (or) write Hindi that well, so asking and listening to audio, I did that.}'' (P12).

Post-surgery limitations on screen use also indicated that speech input and audio output the preferred modality for patients, with 18 out of 26 audio questions by patients asked post-surgery.
31 participants used the related questions at least once. 
As P7 stated:
    ``\textit{I used the audio message once. After that there was no need for me }[to type or use audio]\textit{... I just kept selecting from the }[related questions]\textit{ options.}'' 
While they found the related questions feature convenient and frequently used it, there were instances of dissatisfaction when (in eight cases) the answers to those resulted in ``\textsf{I don't know}'' responses.
This issue arose because the \cataractbot{} system uses an LLM call to suggest three questions based on the users' last question, independent of the knowledge base.
This approach aimed to uncover gaps in the knowledge base by not restricting questions to the curated knowledge base.

On the expert end, they could only interact with \cataractbot{} using text.
This design choice facilitated a ``\textit{hygiene check}'' (D4) for thoroughness and precision in experts' edits, aligning with previous findings~\cite{textmssg-brenna-chi23}.

\imwutpaste{Patients mentioned utilizing the persistent information provided by \cataractbot{} as a reference and sharing it with others undergoing cataract surgery.
Also, as per D1, this information can help when discussing their surgery with experts outside of the \anonymousHospital{} ecosystem.
\imwutdelete{``[Patients]\textit{ can share the same information with their doctor at their native place... So that }[communication with another doctor]\textit{ becomes a little easier.}'' (D1).}
However, it also increased the sense of accountability among experts, making them ``\textit{somewhat nervous}''.
They felt the need to be ``\textit{very lawyer-like in our conversation because we don't want it to come back and bite us tomorrow.}'' (D4).}

\subsubsection{Multilinguality}
\label{findings:features:multilingual}
\imwutdelete{Experts interacted with \cataractbot{} only in English.}
The majority of information seekers (29 out of 49) selected English, while 7 opted for Hindi, 6 Kannada, 5 Tamil, and 2 Telugu.
None of the participants changed their initially selected language.
Given the challenges of typing in Indic languages on smartphones~\cite{swarachakra}, our non-English participants predominantly relied on speech input and selecting from related questions.
In total, patients who chose English typed 35 questions, asked 0 questions in speech, and tapped 6 related questions.
In contrast, patients opting for Indic languages typed 8 questions, sent 26 audio questions and tapped 14 related questions in total.
A similar trend was observed among attendants.
Patients found value in local language interaction \imwutdelete{and speech }as input, helping them use the bot independently: ``\textit{Without even asking me for help, he }(patient)\textit{ was able to use that bot as it supports Tamil.\imwutdelete{ I felt it's very useful.}}'' (A13).

The audio queries in Indic languages were transcribed and translated to English using Azure AI services.
However, this was not highly accurate due to the variability in Indic language dialect.
This is a known issue and prior work~\cite{farmchat-imwut18} relied on human wizards to correct such errors. 
Experts
faced challenges in ``\textit{deciphering}'' such messages since they only received the English transcription.
\imwutadd{In the future, there should be a way to share the original audio query with the expert on demand.}

\imwutcut{\imwutdelete{It is common to} type Indic languages in Latin script (\textit{e.g.,} instead of using Devanagari script for Hindi), occurring in three messages.
\imwutdelete{This was not supported by \cataractbot{} due to current translation technology limitations, and it responded with ``\textsf{You seem to be asking the question in a language different from the selected language...}''.}
However, when a message contained a code-mix of Hindi and English, the bot responded accurately:
\begin{quote}
    \textbf{A8}: \textsf{agar opration k baad pain ho raha to kya karna hai?}\\
    \textbf{Bot}: \textsf{If the patient is experiencing pain post-surgery, they should report to the doctor immediately...}
\end{quote}
Even in the selected language, those with limited literacy struggled to formulate questions (similar to findings in prior research~\cite{farmchat-imwut18}).
For example, ``\textsf{Tomorrow operation When should come No phone call received yet?}'' (P18).
There were 65 such questions, of which 26 were asked using audio in an Indic language.
Despite this, the GPT-4 model's robustness enabled the chatbot to answer most (45) of such ill-formulated questions appropriately.}

\subsubsection{\imwutadd{Support for} Multiple Stakeholders}
\label{findings:rq2:multistakeholder}
\imwutpaste{Patients sent a total of 89 messages (4.7$\pm$5.4 message/patient), while attendants sent 254 messages (8.5$\pm$6.8 message/attendant), significantly more than patients, with t(47)=2.04, p<0.05 (Figure \ref{fig:cataractbot_CHARTS}C).
During our interviews, we identified a few reasons for the patients to minimally use the bot, particularly related to demographics.
Patients were typically older with lower levels of literacy and tech-literacy compared to their attendants
These patients trusted the hospital to provide them with all the required information and did not actively seek information themselves.
For instance:
\begin{quote}
    ``Dad didn't try... I was the one thinking about it [the surgery]. He was like, ``\textit{It's OK! Not everyone needs to bother in the family... (to know) about what will be this, what will be that}''. '' -- A3.
\end{quote}
Additionally, patients were advised to minimize screen use post-surgery, contributing to reduced bot interaction.
Interestingly, patients sometimes offloaded the knowledge-seeking `work' to their active attendants---e.g., ``\textit{I have been using it. I told him} [the patient] \textit{what answers I got. So maybe he did not feel the need} (to use it).'' (A20).
}

\imwutpaste{
Moving on to experts, both the doctors and patient coordinators highlighted the significant role of escalation experts.
Among the 309 verified medical and logistical answers, 62.5\% were verified by the operating doctor/coordinator and the rest 37.5\% by the escalation expert.
Doctors mentioned having 2-3 days per week reserved for surgeries, and on these ``\textit{operating days}'', they rarely checked their phones during work hours, leading to more escalations.
Despite their crucial role, escalation experts interacted differently from operating doctors and coordinators. 
Unlike the latter, who had face-to-face meetings with patients, escalation experts lacked contextual knowledge, limiting them to providing generic responses.
}

\subsubsection{Knowledge Base}
\label{findings:rq2:knowledgebase}
Out of the 91 responses edited by experts, 72 (79.1\%) received approval from the knowledge base expert (D4) to be added to the knowledge base.
For 49.3\% of these, the bot's final response (generated by merging the initial response and expert edit suggestion) underwent additional editing by the knowledge base expert.
These changes were minimal with an average relative Hamming distance of 28.2\%,  and mainly addressed aspects of ``\textit{tone and structure}'', such as preferring active  voice and making responses more generally applicable across patients.
This hints that the bot effectively incorporated experts' edits.
\imwutdelete{Over time, experts noticed improvements in the completeness and accuracy of \cataractbot{}'s generated answers, thus reducing their workload: ``\textit{With time, I think the answers did get better... I am mostly just saying yes.}'' (D3).} 
\imwutadd{Between the first and sixth weeks of the study, the proportion of LLM-generated answers that experts marked as ``accurate and complete'' increased by 9.9\% for medical questions and 22.9\% for logistical questions.
This suggests that as the knowledge base was updated, the experts' verification workload decreased.}

\subsection{\imwutadd{Bot Integration into Doctor-Patient Workflows}}
\label{findings:rq3}

\subsubsection{\imwutadd{Workflows}}
\label{findings:rq3:workflows}
\imwutpaste{The majority of messages (71.4\%) were sent in the 8-day period from 5 days prior to surgery to 2 days post-surgery with the highest number of messages (76) sent on the surgery day (Figure~\ref{fig:cataractbot_CHARTS}B).}
\imwutdelete{156 messages were sent before the surgery day and 111 post the surgery day.
We conducted a one-way ANOVA on the number of messages asked by each patient-attendant group pre-surgery, on the day of surgery, and post-surgery, and found statistical significance with F(2, 60)=3.69, p<.05.
Post-hoc pairwise comparisons with Bonferroni correction indicated a statistically significant difference between `pre' and `on' surgery (t(30)=2.7, p<.05) and between `pre' and `post' surgery (t(30)=3.2, p<.05).
This suggests that patients and attendants are more inquisitive before surgery, potentially due to heightened anxiety, aligning with prior work~\cite{patient-info-seeking-analee-1990}. 
The higher pre-surgery usage may also be due to the novelty effect.
With respect to the types of queries, logistical questions were mainly asked before the surgery, reaching their peak one day before the surgery day (Figure \ref{fig:cataractbot_CHARTS}B), as anticipated.
In contrast, experts expressed that questions on the day of surgery might be unnecessary:
\begin{quote}
    ``Multiple questions were asked on the day of surgery, ``\textit{When is he going to be discharged?}''... You are in the hospital... Why use the bot?!... You can walk 10 steps and ask the staff.'' -- D1.
\end{quote}
}
\imwutpaste{Messages were sent throughout the day (7am to 11pm), as A18 mentioned: ``\textit{Whenever I am idle at work or at home... whatever (question) comes to my mind, I just ask.}''}
\imwutpaste{The workflow employed by information seekers varied.
Most participants would ask a question, leave their phones, and check the answer later when free.
They hardly noticed the wait due to being occupied with their day-to-day activities.
In such cases, notifications played a key role.
The message stating that the expert had provided a verified response (Figure \ref{fig:cataractbot_textmessage}G), made the otherwise unnoticeable green tick (Figure \ref{fig:cataractbot_textmessage}F) prominent.}
Participants, when in need of a quick response, either relied on the initial (LLM-generated) answer or re-verified it on the internet.

\imwutadd{19/22 survey respondents }\imwutpaste{(86.4\%) agreed that the bot responded quickly. 
On average, the bot took 11.8$\pm$10.2 seconds to respond. 
Expert verification took 162.9$\pm$172.3 minutes.
We found verification without edits (151.1$\pm$147.1 minutes) to be significantly faster than verification with edits (191.3$\pm$172.4 minutes) with t(265)=2.7, p<0.05.
Participants were ``\textit{okay to wait}'' for the expert's verification, since the response was being verified by ``\textit{busy doctors}''. This demonstrated their clear understanding of the bot's workflow.
\begin{quote}
    ``I know that it will take its own time... based on their [doctors'] busy schedule, their availability... It's not like there's some dedicated doctor monitoring it all the time, right? If that is the case, then (the) bot is not required. They [doctor] can directly reply.'' -- A17.
\end{quote}
}
Notifications played a key role.
The message stating that the expert had provided a verified response (Figure \ref{fig:cataractbot_textmessage}G), made the otherwise unnoticeable green tick (Figure \ref{fig:cataractbot_textmessage}F) prominent and the bot's process self-explanatory.

\imwutpaste{For experts, the asynchronous nature of the bot provided them the flexibility to offer verifications at their convenience. 
The majority of responses (31) were provided by experts at 4-5 pm, post their busy workday, as most OPD and surgeries finish by 4 pm. 
The second-highest responses (28) occurred at 1-2 pm, during their lunch break.
This usage pattern was described as:
    ``\textit{I use my mobile phone only when I go for breaks... lunch break, coffee break, or in the morning before I start OPD. The rest of the time, my }(cellular)\textit{ data is off.}'' (D1).
Experts refrained from using their phones in front of patients due to the negative perception associated with using phones for personal reasons.
We found that the experts were ``\textit{fine}'' with the additional task of verifying the bot's responses, as articulated by D4:
``\textit{It takes only about 15-20 seconds to answer a question... It was not really eating into my time... It's much easier than replying to my mailbox.}''}

\imwutdelete{Opinions on using the chatbot after work hours varied: two experts were reluctant, while four were in favor of using it ``\textit{anytime}''.}
\imwutpaste{As five experts used \cataractbot{} on their personal WhatsApp accounts, it blurred the boundary between their home and work, similar to previous findings~\cite{whatsapp-boundaries-mols-2021, ding-wechat-chi2020}.
Our experts also noted that WhatsApp had a low signal-to-noise ratio, resulting in instances where they missed \cataractbot{} messages:
``\textit{My WhatsApp has 5000+ }[unread]\textit{ messages... college groups, family groups... I just see messages in the important groups.}'' (D2).
Addressing this issue, the escalation doctor D4 suggested that experts should be recommended to ``\textit{pin}'' \cataractbot{} on WhatsApp, ensuring it remains consistently at the top of their chat list.}
\label{findings:privacy}
\subsubsection{Personalization vs Privacy}
Three participants valued that
the bot could not access their medical records.
\begin{quote}
    ``Using personal health information [by the bot] is... a violation of patients' rights. If their data is added... external agencies can extract that information and might use it for advertising.'' -- A17.
\end{quote}
These participants were educated (holding Master's degrees) and considered personalization as ``\textit{an unnecessary add-on}''.
They raised concerns about complications associated with sharing PII with LLMs, both in terms of legislation and ``\textit{privacy threats}''. 
A1, for example, remarked ``\textit{I would definitely not want my mom's data to be out there, all over the internet.}''.
Moreover, lack of personalization enabled usage of the chatbot for others undergoing cataract surgery: ``\textit{Two people I know are going to have their cataract surgery soon... I'll use this bot for them.}'' (A11).

On the other hand, fueled by the recent hype around AI and LLMs, eight participants expected ``\textit{personalized}'' answers to their queries with little concern for privacy, and were slightly disappointed with generic responses.
This aligns with prior findings indicating minimal privacy concerns related to digital data in the global south, specifically in India~\cite{okolo2021cannot}.
Interestingly, three participants did not even consider their medical data to be ``\textit{private}'', as they stated, ``\textit{It's nothing personal to be hidden. All }(of us)\textit{ have common health issues.}'' (A14).
Moreover, two participants questioned why others would be interested in their medical data, as ``\textit{it is not useful for others}'' (P18).

Two participants wanted personalization, along with the sharing of patient data, to be an ``\textit{opt-in}'' feature where individuals conduct a cost-benefit analysis and make an informed choice based on their preferences.
They believed that a personalized bot would offer more relevant, and fewer ``\textsf{I don't know}'', responses.
\begin{quote}
    ``
    If you don't want to share your information, it (the bot) should say ``\textit{If you ask personal questions, I won't have access, so you may have to visit the hospital.}''... We have to compromise.'' -- A21.
\end{quote}

Five experts voted to integrate the chatbot with patient data, including schedules and consultation history, to offer personalized responses.
D3 suggested integrating
patient medical history to facilitate customized responses from the bot and doctors: ``\textit{We don't even know whether they're diabetic, they're hypertensive. So, when they ask ``\textit{What should I eat and come?}''... I am not sure what to say.}'' (D3).\imwutdelete{~Although the onboarding form attempts to capture such information in the `Extra Details' field, none of our participants provided that information due to time constraints.}
However, D1 and D2 also voiced skepticism towards the LLMs capability to understand complex patient records accurately.
``\textit{There are a lot of variables... the surgeon, the  patient, their tests... It's too much information to process.}'' (D1).
In addition, the knowledge base expert cautioned that highly personalized responses would result in minimal updates to the knowledge base, potentially increasing \cataractbot{}-related workload for experts and negatively impacting chatbot adoption among experts.

\section{Discussion}
\label{discussion}
In this paper, we present \cataractbot{}, an LLM-powered experts-in-the-loop WhatsApp chatbot to address the information needs of patients undergoing cataract surgery. 
\imwutadd{Building on limited prior research in multi-stakeholder settings~\cite{yang2023talk2care,jo2023understanding}, we provide valuable insights into LLM-mediated interactions among patients, attendants, doctors, and coordinators.}
Our field study
revealed positive evidence of \cataractbot{}'s usefulness and usability across stakeholders. 
Doctor-verified responses were key to patients' and attendants' trust in and engagement with the bot.
They appreciated that \cataractbot{} instantly answered questions anytime, saving time by reducing the need for hospital calls or visits.
Those with limited (tech-)literacy found it easy to use because of the multilingual and multimodal support. 
Doctors and coordinators commended \cataractbot{} for acting as a facilitator, creating a layer of privacy between them and patients, and providing them the flexibility to verify responses at their convenience.

\imwutcut{We designed and evaluated our experts-in-the-loop chatbot system in the context of cataract surgery at \anonymousHospital{} in India
Our open-sourced framework can also be adapted beyond cataract surgery, to accompany more complex, atypical, or long-term treatments--e.g., glaucoma surgery, cancer treatment, or pregnancy--where easy access to verified information would be valuable for patients.
The applicability can be extended further, to fields beyond the medical domain, such as law, finance and education.
In these scenarios, end-users (e.g., defendants, taxpayers, or students) could receive synchronous responses from an LLM using a custom knowledge base, and subsequent asynchronous verified responses from experts (e.g., lawyers, accountants, or teachers).
Expert-mediated bots can even be used to upskill less-trained workforce members.
For instance, these bots could serve as valuable learning resources for community health workers, who typically have minimal medical training.\imwutdelete{  The resource could boost their confidence, while helping them provide accurate information to their care recipients.}}

Below, we \imwutdeleteagain{identify design implications, raising }\imwutaddagain{raise} critical insights and open questions that warrant further attention for such experts-in-the-loop chatbots to succeed on a larger scale across domains.


\imwutdeleteagain{\subsection*{7.1 Design Implications}}

\subsection{Scaling and Generalization}
\label{discussion:designimplications:scaling}
\imwutpaste{We designed and evaluated our experts-in-the-loop chatbot system in the context of cataract surgery in India.}
\imwutadd{
Expanding it beyond India may require further research.
In the United States, for example, insurance plays a crucial role in healthcare. 
Patients often require prior authorization for treatments, and they would value a bot providing information in navigating these processes~\cite{sun2013patientproviderportal}. 
This necessitates involving insurance-specific experts in knowledge base creation and response verification.
Also, our study raised concerns about expert accountability due to the persistence of written communication in the bot.
This issue could be more pronounced in the Global North, where doctors are increasingly wary of lawsuit risks~\cite{sun2013patientproviderportal}.
Balancing these needs and tensions will be critical for \cataractbot{}'s adoption across geographies.
Finally, health communication is most effective when users perceive the source as demographically and attitudinally similar to themselves, not just contextually relevant~\cite{kreuter2004cultureinhealth}.
This highlights the need to adapt \cataractbot{} for different settings by (1) parametrizing context-specific details, such as payment methods and hospital protocols, and (2) tailoring language to ensure cultural relevance through familiar terms and metaphors.}

\imwutpaste{
Our open-sourced framework can be adapted beyond cataract surgery to support more complex, atypical, or long-term treatments--e.g., glaucoma surgery, cancer treatment, or pregnancy--where easy access to verified information would be valuable for patients}, \imwutaddagain{improving their health literacy and enabling better self-care.
While conducting such studies, aligning the long-term research agenda with the operational priorities of busy hospitals could pose challenges, as cautioned by \citet{agapie2024researchhcihealthsupportingteams}. 
It is crucial to ensure ``\textit{that clinicians are fully aware of the motivations and methodologies of the} (research) \textit{process, which is very different from a normal clinical situation}''~\cite{newell2000usersensitiveinclusivedesign}.
For instance, during our study, securing trust, recruiting hospital staff, and minimizing administrative delays required active endorsement from a senior doctor.
This highlights the need to thoughtfully leverage existing authority structures in institutional settings.
We also observed that positioning the technology as a long-term solution to reduce workload helped minimize clinicians' resistance during the initial onboarding phase.}

Beyond the medical domain, the applicability of our framework can be extended further, to fields such as law, finance and education.
In these scenarios, end-users (e.g., defendants, taxpayers, or students) could receive synchronous responses from an LLM using a custom knowledge base, and subsequent asynchronous verified responses from experts (e.g., lawyers, accountants, or teachers).
Expert-mediated bots can even be used to upskill less-trained workforce members.
For instance, these bots could serve as valuable learning resources for community health workers, who typically have minimal medical training\imwutadd{~\cite{yadav-feeding-2019}}.

\subsection{Expert-in-the-loop Approaches}
\label{discussion:healthcarechatbots:expertinloop}
Prior studies have raised concerns about the application of LLMs in healthcare scenarios, due to their inconsistency, potential errors, and bias\imwutadd{~\cite{denecke2024potentialofllms, au2023ai}}.
Moreover, users who are less (technologically) literate face a heightened risk of harm~\cite{kapania2022attitude}.
Our experts-in-the-loop framework not only addresses these concerns but also adheres to OpenAI's usage policies~\cite{openai_usage_policies}, which state that ``\textit{tailored medical/health advice cannot be provided without review by a qualified professional}''.
By using our framework, chatbots can operate at the ``\textit{sweet spot of patient-LLM-clinician collaboration}''~\cite{hao2024LLMapplicationsinpatienteducation}, where the capabilities of technology and human judgment are integrated.
In our study, this ensured reliability and cultivated trust, as information seekers knew that their operating doctor was verifying responses.
However, trust in the initial automated answers reduced when these answers were subsequently marked as incorrect/incomplete by the doctor.
To address this in future work, one design approach could be to eliminate the initial automated response, providing only expert-verified answers. 
We caution that this may lead to long wait times, potentially discouraging use. 
For high-risk use cases like cancer treatment,
verified accuracy may be prioritized over immediacy.
Future research could compare the two paradigms:
a hybrid system with synchronous automated answers followed by asynchronous verification versus a fully asynchronous system delivering only verified answers.
Such a study would provide valuable insights into balancing immediate access with the accuracy required for critical medical information in chatbot design.
\imwutadd{Doctors are a scare resource in healthcare ecosystems. 
Our chatbot system can help scale their impact, by reducing the time spent answering routine questions at length during consultations and calls, thus freeing up time for additional patient care. However, as doctors had to verify responses
during deployment, 
\cataractbot{} also added to their workload.
To minimize this, one approach is to deliver pre-verified answers to specific questions that the expert has repeatedly marked as correct.
Given the potential for LLM hallucinations~\cite{denecke2024potentialofllms}, cached responses must be carefully matched to similar patient demographics and question phrasing before reuse.
\imwutaddagain{Moreover, the information seeker should retain agency within this process.
If a pre-verified answer is deemed insufficient or irrelevant, they should be able to flag it to initiate the expert-in-the-loop workflow for a revised, verified response.}
Another approach is to use a second LLM for quality control while generating each response, as
GPT-4's evaluations of AI-generated responses align well with human expert assessments~\cite{gumma2024healthparikshaassessingragmodels}.
This maker-checker verification module would enable tracking of the bot's response quality, especially as the knowledge base expands.
The module could also evaluate the confidence level in the generated response, or the perceived urgency of the question, which could be used to prioritize sending nudges to experts for verification.}
\subsection{Multimodality and Personalization}
\label{discussion:designimplications:multimodalityandpersonalization}
Our system faced challenges in 
accurately interpreting questions from voice messages due to limitations in transcription and translation technologies.
The same inaccurate transcriptions were shared with doctors, leading to confusion.
\imwutadd{Such issues are less likely when deploying a voice-based input modality to a predominantly English-speaking population or in tier-1 language settings in the Global North, as seen in prior work~\cite{yang2023talk2care}.}
\imwutaddagain{For future deployments of LLM-powered bots, particularly in low-resource languages, we propose the following enhancements:
(1) implement a dictionary of common errors across languages to serve as a look-up resource, 
(2) provide the LLM with both the original query and its English translation to improve answer generation, and
(3) display a transcription of any audio question back to the information seeker, allowing them to verify or clarify what the bot understood.
These would serve as additional safeguards in cases where language technologies fail.}
Additionally, non-verbal cues in audio messages--such as signs of anxiety, impatience, or the presence of others--were lost in textual transcriptions.
This hindered experts' ability to deliver appropriate care (also noted in \cite{textmssg-brenna-chi23}). 
To address this, we suggest making the original audio clips available to experts on demand.

\imwutdelete{Recognizing that having to listen to each audio message would be added workload for experts, the textual transcription should be shared upfront, with the option to access the audio on-demand.~}

Beyond audio, recent LLM models like GPT-4V offer visual capabilities that could enable such bots to process image-based inputs.\imwutdeleteagain{, aligning with recommendations in prior studies~\cite{farmchat-imwut18}.}
\imwutaddagain{This aligns with the recommendations of \citet{migrant-worker-chatbots-tseng-2023}, suggesting that healthcare information can often be communicated more effectively through images and videos than through text alone.}
For example, a patient could upload a photo of a specific medicine to ask questions about it, instead of providing a written description.
While this could improve the user experience for information seekers, it may increase the workload for experts.
Further, prior research~\cite{hao2024LLMapplicationsinpatienteducation} cautions that although LLMs are proficient in text processing, their ability to handle other forms of media, such as image and video, remains less reliable.


Moving on to personalization, \imwutaddagain{our current \cataractbot{} deployment integrates minimal sensitive data, such as the patient's demographic details and date of surgery.
Going forward, }while tailored responses can boost adoption and utility by providing more relevant information, they also raise privacy concerns, potentially reducing usage among certain users.
\imwutadd{This issue may be more pronounced in the developing world, where there is more awareness around privacy.}
Additionally, experts may face increased workloads as they reference user-specific information for verification, as many personalized edits may not be suitable for inclusion in the generic knowledge base.
Low (tech-)literate users might require assistance in making informed decisions about the use of their data~\cite{okolo2021cannot}.
A binary approach of either sharing all records or none may not address individual preferences.
Instead, users should have the agency to selectively share records they deem relevant and acceptable for the bot's access.
This aligns with the philosophy of India's National Digital Health Mission~\cite{ndhm2020}. 
One solution could be to allow users to upload relevant documents 
or manually enter specific details they want the bot to access.
This can empower users to control shared information while benefiting from a personalized experience.
\imwutadd{Further, when developing \cataractbot{} in May 2023, we chose to use GPT-4, as it was the leading LLM~\cite{gpt4_report}.
However, recent open-source models like LLaMA have shown comparable performance to GPT in medical applications~\cite{li2025gptvsllama}.
Incorporating these models into the \cataractbot{} system would enable local processing of patient data, helping address privacy concerns.}

\subsection{Limitations}
\label{discussion:limitations}
We acknowledge several limitations of this work.
First, results should be interpreted with caution, given the specificities of the study, including \imwuthighlight{a relatively small sample size}, users' first encounter with a medical chatbot, and the cataract treatment scenario.
Positive user responses, minimal increase in expert workload, and other findings should be validated in future studies, preferably with longitudinal designs. 
Second, system-level analyses, such as scalability testing and power measurement, are needed before deploying the proposed solution more broadly.
Finally, although participants reported high trust, repeated ``\textsf{I don't know}'' responses
or inaccurate information during regular use may erode this trust.
We note the potential risk of disseminating inaccurate medical information at scale with chatbot and LLM technologies. 
Therefore, stronger methods are needed to develop, review, and update the knowledge base, as well as to routinely evaluate the accuracy of bot's responses.
\section{Conclusion}
\label{conclusion}
Building a system that addresses patients' information needs by providing expert-verified information is an open problem in healthcare.
We propose a novel solution---an LLM-powered experts-in-the-loop chatbot framework, that utilizes retrieval-augmented generation over a custom knowledge base to provide synchronous responses, and utilizes experts to provide verified responses asynchronously.
Our in-the-wild study involving 49 information seekers (patients and attendants) and 6 experts (4 doctors and 2 coordinators) demonstrated the positive impact of this technological intervention.
Patients not only trusted the chatbot's response, but were also willing to wait, understanding that their busy doctors were verifying them.
Simultaneously, doctors and coordinators, despite their hectic schedules, made time during breaks and at home to help patients by verifying LLM-generated responses.
The favorable feedback from various stakeholders in the healthcare ecosystem indicates that a chatbot, \imwuthighlight{delivered through ubiquitous smartphones and WhatsApp,} can effectively enhance information access in critical healthcare settings, even for individuals with limited literacy and technology experience.

\begin{acks}
Many thanks to all the participants for their time and patience.
\end{acks}

\bibliographystyle{ACM-Reference-Format}
\bibliography{main}

\appendix
\section{Appendix}

\subsection{\cataractbot{} Implementation Details}

\subsubsection{\cataractbot{} LLM Prompts}
\label{appendix:prompt}

The \cataractbot{} system leverages LLM (GPT-4 in our case) for these four tasks (Table \ref{tab:llm-prompts}):


\begin{enumerate}
\item \textbf{Response Generation}: 
For every medical and logistical question asked by the patient/attendant, the system performs a vector search on the Knowledge Base (KB) to extract the three most relevant data chunks related to the query. The LLM is then prompted (Table~\ref{tab:llm-prompts}) to extract an answer for the query from these data chunks. Note: Question Classification is part of the same LLM call to improve efficiency.
\item \textbf{Related Questions Generation}: 
For every medical and logistical question, the system prompts LLM to generate three related questions based on the preceding query. Note: The 72 character limit is due to the WhatsApp's message limit in interactive suggestions.
\item \textbf{Final Response Generation}: If the expert marks the initial LLM-generated response as incorrect or incomplete, the system prompts LLM to generate the final response by merging both the initial response and the expert's edit suggestion.
\item \textbf{Shorten Response}: If the generated response exceeds WhatsApp's message limit of 700 characters, the system prompts LLM to summarize it within the specified character limit.
\end{enumerate}

\begin{table*}[]
\centering
\small
\caption{LLM Prompts used in CataractBot.}
\label{tab:llm-prompts}
\resizebox{\textwidth}{!}{%
\begin{tabular}{l|l|l}
\hline
 &
  \multicolumn{1}{c|}{\textbf{System Prompt}} &
  \multicolumn{1}{c}{\textbf{Query Prompt}} \\ \hline
\textbf{\begin{tabular}[c]{@{}l@{}}Response \\ Generation\end{tabular}} &
  \begin{tabular}[c]{@{}l@{}}You are a Cataract chatbot whose primary goal\\ is to help patients undergoing or undergone a \\ cataract surgery. If the query can be truthfully \\ and factually answered using the knowledge \\ base only, answer it concisely in a polite and \\ professional way. If not, then just say ``I do \\ not know the answer to your question. If this \\ needs to be answered by a doctor, please \\ schedule a consultation.''\\ \\ In case of a conflict between raw knowledge \\ base and new knowledge base, prefer the new \\ knowledge base. One exception to the above \\ is if the query is a greeting or an acknow-\\ ledgement or gratitude. If the query is a \\ greeting, then respond with a greeting. If the \\ query is an acknowledgement or gratitude to\\ the bot's response, then respond with an \\ acknowledgement of the same. Some exam-\\ ples of acknowledgement or gratitude to the \\ bot's response are ``Thank You'', ``Got it'' \\ and ``I understand''. In addition to the above,\\ indicate like a 3-class classifier if the query \\ is ``medical'', ``logistical'' or ``small-talk''. \\ Here, ``small-talk'' is defined as a query \\ which is a greeting or an acknowledgement \\ or gratitude. Answer it in the following\\ json format:\end{tabular} &
  \begin{tabular}[c]{@{}l@{}}The following knowledge base have been \\ provided to you as reference:\\     Raw documents are as follows:\\         \textless{}relevant chunks string\textgreater\\     New documents are as follows:\\         \textless{}relevant updated chunks string\textgreater\\     The most recent conversations are here:\\         \textless{}conversation string\textgreater\\     You are asked the following query:\\         \textless{}user query\textgreater\\ \\ Ensure that the query type belongs to only the \\ above mentioned three categories. When not \\ sure, choose one of ``medical'' or ``logistical''.\end{tabular} \\ \hline
\textbf{\begin{tabular}[c]{@{}l@{}}Related Questions \\ Generation\end{tabular}} &
  \begin{tabular}[c]{@{}l@{}}What are three possible follow-up questions\\  the patient might ask? Respond with the \\ questions only in a python list of strings. Each \\ question should not exceed 72 characters.\end{tabular} &
  \begin{tabular}[c]{@{}l@{}}A patient asked the following query:\\ \textless{}query\textgreater\\ A chatbot answered the following:\\ \textless{}response\textgreater{}\end{tabular} \\ \hline
\textbf{\begin{tabular}[c]{@{}l@{}}Final Response \\ Generation\end{tabular}} &
  \begin{tabular}[c]{@{}l@{}}You are a Cataract chatbot whose primary \\ goal is to help patients undergoing or under-\\ gone a cataract surgery. A cataract patient \\ asks a query and a cataract chatbot answers\\ it. But, the doctor gives a correction to the \\ chatbot's response. Update the cataract \\ chatbot's response by taking the doctor's \\ correction into account. Respond only with\\ the final updated response.\end{tabular} &
  \begin{tabular}[c]{@{}l@{}}A cataract patient asked the following query:\\ \textless{}query\textgreater\\ A cataract chatbot answered the following:\\ \textless{}response\textgreater\\ A doctor corrected the response as follows:\\ \textless{}correction\textgreater{}\end{tabular} \\ \hline
\textbf{\begin{tabular}[c]{@{}l@{}}Shorten \\ Response\end{tabular}} &
  \begin{tabular}[c]{@{}l@{}}You are a Cataract chatbot, and you have \\ to summarize the answer provided by a bot. \\ Please summarise the answer in 700 char-\\ acters or less. Only return the summarized \\ answer and nothing else.\end{tabular} &
  \begin{tabular}[c]{@{}l@{}}You are given the following response: \\ \textless{}response\textgreater{}\end{tabular} \\ \hline
\end{tabular}
}
\end{table*}

\imwutpaste{\subsubsection{\cataractbot{}'s Support for Multimodal Communication}
\label{appendix:multimodal}
\begin{figure*}
  \includegraphics[width=\textwidth]{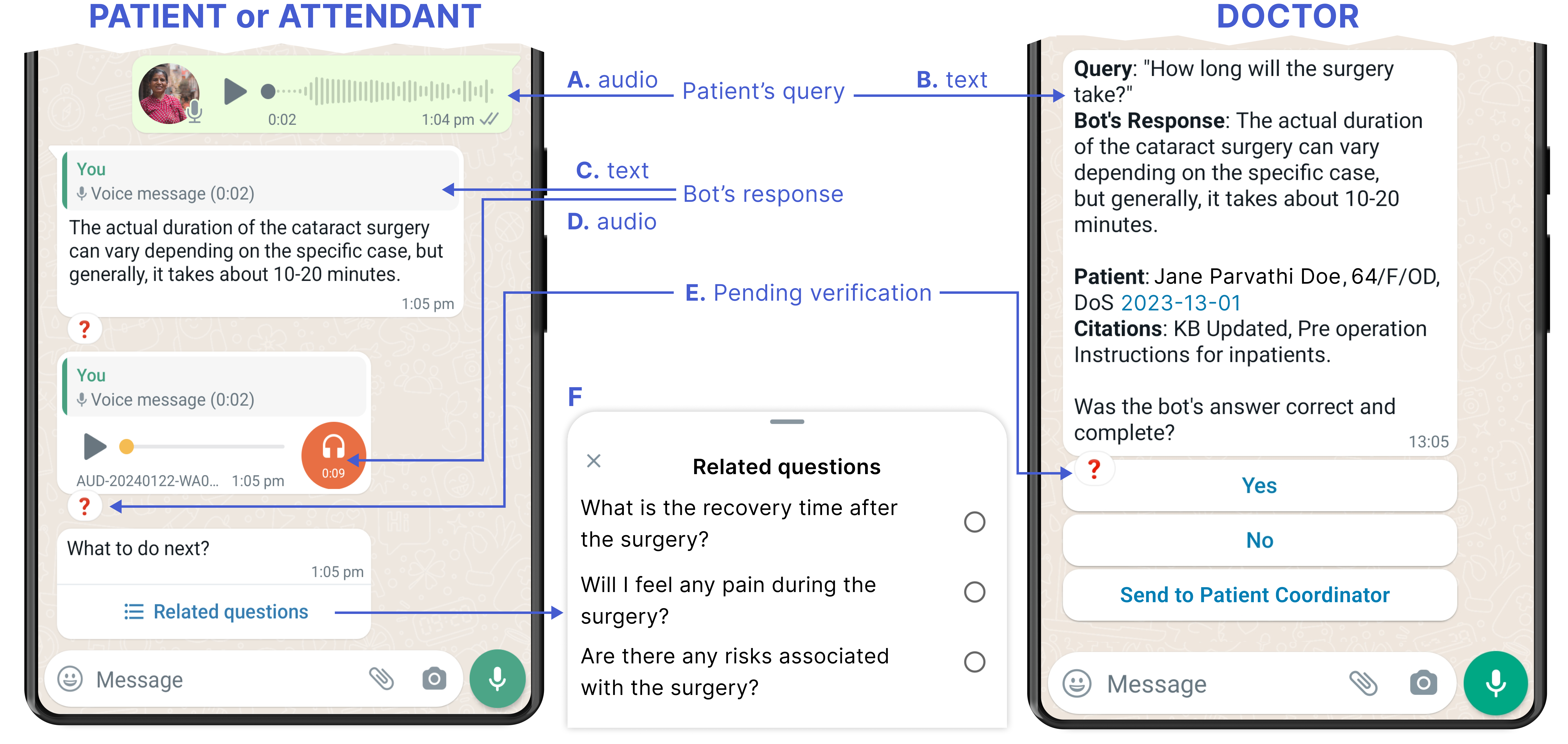}
  \caption{\imwutpaste{A question asked (using audio), receiving an unverified response and Related Questions from \cataractbot{}.}}
  \Description{This image displays a split-screen interface showing the communication between a patient/attendant (left) and doctor (right) using CataractBot. The patient initially asks "How long will the surgery take?" via an audio message (labeled A), which is transcribed to text (B) for the doctor's view. The bot provides a response (C): "The actual duration of the cataract surgery can vary depending on the specific case, but generally, it takes about 10-20 minutes," which is delivered to the patient as both text and audio (D). The interface indicates this response is pending verification (E) from the doctor, who also has access to patient Jane Parvathi Doe's details (64/F/OD with surgery date 2023-12-01). Below this exchange, the system displays related questions (F) that the patient might want to ask next, including recovery time, potential pain during surgery, and associated risks. On the doctor's side, they can verify whether the bot's answer is correct and complete by selecting "Yes" or "No," or send the query to a Patient Coordinator.}
  \label{fig:cataractbot_audiomessage}
\end{figure*}
See Figure \ref{fig:cataractbot_audiomessage}.}

\subsection{Participant Demography}
\label{appendix:participant demography}
\begin{table*}[]
\centering
\caption{Demographic details for 19 patients and 30 attendees, with their interview participation, the duration of \cataractbot{} usage (calculated as the difference between the first and last day of messages sent), and the total number of messages sent.}
\label{tab:demography-patients}
\resizebox{\textwidth}{!}{%
\begin{tabular}{ccccc|ccccc|cccc}
\hline
\textbf{PId} &
  \textbf{Age} &
  \textbf{Sex} &
  \textbf{Language} &
  \textbf{\begin{tabular}[c]{@{}c@{}}Highest\\ Education\end{tabular}} &
  \textbf{AId} &
  \textbf{Age} &
  \textbf{Sex} &
  \textbf{Language} &
  \textbf{\begin{tabular}[c]{@{}c@{}}Highest\\ Education\end{tabular}} &
  \textbf{\begin{tabular}[c]{@{}c@{}}Surgery-day\\ Interview\end{tabular}} &
  \textbf{\begin{tabular}[c]{@{}c@{}}Post-surgery\\ Interview\end{tabular}} &
  \textbf{Days} &
  \textbf{\begin{tabular}[c]{@{}c@{}}\# of \\ Messages\end{tabular}} \\ \hline
  &         &          &    &  & \textbf{A1}  & 43 & M & English & Masters   & Yes & Yes & 7  & 9  \\
             & \textbf{} & \textbf{} & \textbf{} & \textbf{} & \textbf{A2}  & 31 & F & Kannada & Grade 12  & Yes & Yes & 3  & 13 \\
\textbf{P3}  & 56        & M         & Tamil     & $\leq$Grade 10 & \textbf{A3}  & 35 & M & English & Bachelors & Yes & No  & 4  & 6  \\
\textbf{P4}  & 52        & M         & English   & Grade 12  & \textbf{A4}  & 39 & F & English & Grade 12  & Yes & No  & 4  & 8  \\
\textbf{P5}  & 67        & M         & Kannada   & $\leq$Grade 10 & \textbf{A5}  & 40 & M & Kannada & Bachelors & Yes & No  & 5  & 4  \\
             & \textbf{} & \textbf{} & \textbf{} & \textbf{} & \textbf{A6}  & 35 & M & Telugu  & $\leq$Grade 10 & Yes & No  & 1  & 2  \\
\textbf{P7}  & 58        & M         & Telugu    & $\leq$Grade 10 & \textbf{A7}  & 25 & M & English & Masters   & Yes & Yes & 23 & 17 \\
             & \textbf{} & \textbf{} & \textbf{} & \textbf{} & \textbf{A8}  & 39 & M & Hindi   & Bachelors & Yes & No  & 6  & 6  \\
\textbf{P9}  & 69        & M         & English   & Masters   & \textbf{A9}  & 38 & F & English & Masters   & Yes & Yes & 6  & 14 \\
             & \textbf{} & \textbf{} & \textbf{} & \textbf{} & \textbf{A10} & 71 & M & English & Bachelors & Yes & Yes & 26 & 15 \\
             & \textbf{} & \textbf{} & \textbf{} & \textbf{} & \textbf{A11} & 32 & M & Hindi   & $\leq$Grade 10 & Yes & No  & 14 & 6  \\
\textbf{P12} & 65        & M         & Hindi     & Bachelors & \textbf{A12} & 31 & F & English & Masters   & Yes & Yes & 2  & 6  \\
\textbf{P13} & 69        & M         & Tamil     & Grade 12  & \textbf{A13} & 35 & M & English & Masters   & Yes & Yes & 25 & 18 \\
\textbf{P14} & 56        & F         & Kannada   & $\leq$Grade 10 & \textbf{A14} & 33 & F & English & Bachelors & Yes & Yes & 22 & 30 \\
             & \textbf{} & \textbf{} & \textbf{} & \textbf{} & \textbf{A15} & 57 & M & English & Grade 12  & Yes & No  & 4  & 4  \\
\textbf{P16} & 57        & F         & Hindi     & $\leq$Grade 10 & \textbf{A16} & 25 & M & English & Masters   & Yes & No  & 1  & 1  \\
             & \textbf{} & \textbf{} & \textbf{} & \textbf{} & \textbf{A17} & 39 & M & English & Bachelors & Yes & Yes & 2  & 8  \\
\textbf{P18} & 56        & M         & English   & Grade 12  & \textbf{A18} & 23 & M & English & Masters   & Yes & Yes & 27 & 24 \\
\textbf{P19} & 41        & M         & English   & Masters   & \textbf{A19} & 36 & F & English & Bachelors & Yes & No  & 19 & 15 \\
\textbf{P20} & 64        & M         & Tamil     & Grade 12  & \textbf{A20} & 60 & F & Tamil   & Grade 12  & Yes & No  & 4  & 16 \\
             & \textbf{} & \textbf{} & \textbf{} & \textbf{} & \textbf{A21} & 43 & M & English & Masters   & Yes & No  & 15 & 19 \\
             & \textbf{} & \textbf{} & \textbf{} & \textbf{} & \textbf{A22} & 35 & F & English & Bachelors & Yes & No  & 7  & 11 \\
\textbf{P23} & 54        & M         & English   & Bachelors & \textbf{}    &    &   &         &           &     &     & 7  & 8  \\
\textbf{P24} & 66        & M         & Kannada   & Bachelors & \textbf{A24} & 37 & M & English & Masters   & No  & No  & 1  & 1  \\
             & \textbf{} & \textbf{} & \textbf{} & \textbf{} & \textbf{A25} & 40 & F & Tamil   & $\leq$Grade 10 & No  & No  & 7  & 3  \\
\textbf{P26} & 65        & M         & Hindi     & Bachelors & \textbf{A26} & 34 & M & English & Bachelors & No  & No  & 19 & 15 \\
\textbf{P27} & 64        & M         & Hindi     & Bachelors & \textbf{A27} & 35 & M & English & Bachelors & No  & No  & 18 & 14 \\
             & \textbf{} & \textbf{} & \textbf{} & \textbf{} & \textbf{A28} & 40 & M & English & Masters   & No  & No  & 1  & 1  \\
\textbf{P29} & 45        & F         & Kannada   & $\leq$Grade 10 & \textbf{A29} & 29 & F & English & Bachelors & No  & No  & 15 & 10 \\
\textbf{P30} & 50        & M         & Hindi     & Masters   & \textbf{A30} & 25 & F & English & Bachelors & No  & No  & 5  & 3  \\
\textbf{P31} & 63        & M         & Kannada   & Bachelors & \textbf{A31} & 32 & M & English & Bachelors & No  & No  & 15 & 36 \\ \hline
\end{tabular}%
}
\end{table*}
\begin{table}[]
\imwutadd{
\centering
\caption{\imwutadd{Demography details of doctors and patient coordinators who participated in the formative and/or deployment study}}
\label{tab:demography-experts}
\resizebox{\textwidth}{!}{%
\begin{tabular}{cc|ccccc|cc}
\hline
\textbf{Id} &
  \textbf{Role(s)} &
  \textbf{Age} &
  \textbf{Sex} &
  \textbf{\begin{tabular}[c]{@{}c@{}}Highest\\ Education\end{tabular}} &
  \textbf{Experience} &
  \textbf{\begin{tabular}[c]{@{}c@{}}Surgeries\\ per week\end{tabular}} &
  \textbf{\begin{tabular}[c]{@{}c@{}}Formative\\ study?\end{tabular}} &
  \textbf{\begin{tabular}[c]{@{}c@{}}Deployment\\ study?\end{tabular}} \\ \hline
\textbf{D1} & Operating doctor                         & 42 & F & Masters   & 10+ years & 10-20      & No  & Yes \\
\textbf{D2} & Operating doctor                         & 44 & F & Masters   & 15+ years & 10-20      & No  & Yes \\
\textbf{D3} & Operating doctor                         & 46 & M & Masters   & 20+ years & 30-40 & Yes & Yes \\
\textbf{D4} & Escalation doctor, Knowledge base expert & 46 & M & Masters   & 20+ years & 15-20 & Yes & Yes \\
\textbf{D5} & Doctor                                   & 30 & M & Masters   & 6+ years  & 10    & Yes & No  \\
\textbf{D6} & Doctor                                   & 29 & F & Masters   & 2+ years  & 1-2   & Yes & No  \\ \hline
\textbf{C1} & Operating patient coordinator            & 33 & F & Bachelors & 10+ years & N/A   & Yes & Yes \\
\textbf{C2} & Escalation patient coordinator           & 36 & F & Bachelors & 15+ years & N/A   & No  & Yes \\
\textbf{C3} & Patient coordinator                      & 45 & F & Bachelors & 10+ years & N/A   & Yes & No  \\ \hline
\end{tabular}%
}}
\end{table}

See Tables~\ref{tab:demography-patients} and \ref{tab:demography-experts}.

\subsection{Chatbot Usability Evaluation Form}
\label{appendix:usabilityform}
Responses were given on a 5-point Likert scale, ranging from \textit{Strongly Disagree} to \textit{Strongly Agree}.
\begin{enumerate}
    \item CataractBot understands me well.
    \item CataractBot's responses are easy to understand.
    \item CataractBot's responses were useful, appropriate, and informative.
    \item CataractBot responds quickly.
    \item CataractBot seems to have a good grasp of medical knowledge.
    \item CataractBot is kind and helpful.
    \item CataractBot is easy to use.
    \item I would be willing to use CataractBot (or a similar bot before/after a major surgical treatment) in future.
\end{enumerate}

\imwutadd{\subsection{Summary of Key Findings}
\label{appendix:findings-summary}
\begin{table}[]
\imwutadd{
\centering
\caption{\imwutadd{Key findings from CataractBot deployment study}}
\label{tab:findings-summary}
\resizebox{\textwidth}{!}{%
\begin{tabular}{l|l|l|l}
\hline
\textbf{Research Q} &
  \textbf{Theme} &
  \textbf{Finding (Information Seekers)} &
  \textbf{Finding (Experts)} \\ \hline
\multirow{2}{*}{\begin{tabular}[c]{@{}l@{}}RQ1: \\ Information \\ Needs\end{tabular}} &
  Reasons for usage &
  \begin{tabular}[c]{@{}l@{}}
  • Asking forgotten or uncomfortable questions\\ 
  • Clarifying and verifying information\\ 
  • Seeking updates\end{tabular} &
  • Commitment to patients \\ \cline{2-4} 
 &
  \begin{tabular}[c]{@{}l@{}}Reasons for\\ lack of usage\end{tabular} &
  \begin{tabular}[c]{@{}l@{}}
  • In-person expert interactions suffice\\
  • Lack of questions\\ 
  • ``I don't know'' responses\end{tabular} &
  \begin{tabular}[c]{@{}l@{}}
  • Persistent medium gives rise to\\ 
  accountability concerns\end{tabular} \\ \hline
\multirow{5}{*}{\begin{tabular}[c]{@{}l@{}}RQ2: \\ Features\end{tabular}} &
  LLM-powered &
  \begin{tabular}[c]{@{}l@{}}
  • Instant answers.\\ 
  • Saves time.\\ 
  • Understands complex/ill formed questions\end{tabular} &
  \begin{tabular}[c]{@{}l@{}}
  • Accurate answers\\ 
  • (Informal) corrections are quick and easy\end{tabular} \\ \cline{2-4} 
 &
  Experts-in-the-loop &
  \begin{tabular}[c]{@{}l@{}}
  • Drives trust and engagement\\ 
  • Lowers perceived intelligence of bot\end{tabular} &
  \begin{tabular}[c]{@{}l@{}}
  • Adds a layer of privacy from patients\\ 
  • Corrections expand bot's answers\\
  • No access to final answer reduces both \\workload and transparency\end{tabular} \\ \cline{2-4} 
 &
  \begin{tabular}[c]{@{}l@{}}Multimodality and\\ multilinguality\end{tabular} &
  \begin{tabular}[c]{@{}l@{}}
  • Independent usage by older, \\ less educated, visually impaired patients\end{tabular} &  
  • Hard to decipher English transcriptions \\ \cline{2-4} 
 &
  \begin{tabular}[c]{@{}l@{}}Support for \\ multiple stakeholders\end{tabular} &
  \begin{tabular}[c]{@{}l@{}}
  • Attendants (younger, educated) message \\ more than patients\end{tabular} &
  \begin{tabular}[c]{@{}l@{}}
  • Escalation experts are crucial, but \\ lack patient context\end{tabular} \\ \cline{2-4} 
 &
  Knowledge base &
   &
  \begin{tabular}[c]{@{}l@{}}
  • Minimal edits after first expert verification\\
  • Bot's performance improves over time \\ 
  • Experts' workload eases\end{tabular} \\ \hline
\multirow{2}{*}{\begin{tabular}[c]{@{}l@{}}RQ3: \\ Workflow\end{tabular}} &
  Workflows &
  \begin{tabular}[c]{@{}l@{}}
  • Messages spread throughout hours of day\\ 
  • Most messages on surgery day\\ 
  • Waiting for verification is fine\end{tabular} &
  \begin{tabular}[c]{@{}l@{}}
  • Verify at their convenience\\ 
  • WhatsApp blurs work-home boundaries\end{tabular} \\ \cline{2-4} 
 &
  \begin{tabular}[c]{@{}l@{}}Personalization \\ vs. privacy\end{tabular} &
  \begin{tabular}[c]{@{}l@{}}
  • No personalization implies limited usefulness\\ 
  • Medical history not seen as private by some\\
  • Personalization wanted on opt-in basis\end{tabular} &
  \begin{tabular}[c]{@{}l@{}}
  • Lack of medical history makes verification\\ challenging\\ 
  • Personalization would add complexity\\ and workload\end{tabular} \\ \hline
\end{tabular}%
}}
\end{table}
See Table \ref{tab:findings-summary}}

\end{document}